\documentclass[aps,prd,nofootinbib,twocolumn,superscriptaddress,preprintnumbers,letterpaper,natbib,nofootinbib]{revtex4-2}
\pdfoutput = 1
\usepackage{amssymb}
\usepackage{amsmath}
\usepackage{epsfig}
\usepackage[usenames,dvipsnames]{color}
\usepackage{hyperref}
\usepackage{comment}
\usepackage{color}
\usepackage{cleveref}
\usepackage{multirow}
\usepackage{capt-of}
\usepackage{siunitx}
\usepackage{graphicx}
\usepackage{gensymb}
\usepackage[caption=false]{subfig}
\usepackage{slashed}
\usepackage{tabularx}
\usepackage{enumitem}
\usepackage{multirow}
\usepackage{mathtools}
\usepackage{soul}

\usepackage[page,toc,titletoc,title]{appendix}

\hypersetup{
    colorlinks=true,       % false: boxed links; true: colored links
    linkcolor=Cyan,          % color of internal links (change box color with linkbordercolor)
    citecolor=Magenta}

\makeatletter

\begin{document}

\title{Axion Star Explosions: A New Source for Axion Indirect Detection}
\author{Miguel Escudero}
\email{miguel.escudero@cern.ch}\thanks{-- \href{https://orcid.org/0000-0002-4487-8742}{0000-0002-4487-8742}}
\affiliation{Theoretical Physics Department, CERN, 1211 Geneva 23, Switzerland}
\author{Charis Kaur Pooni}
\email{charis.pooni@kcl.ac.uk}\thanks{-- \href{https://orcid.org/0000-0002-6658-3478}{0000-0002-6658-3478}}
\affiliation{Physics, King's College London, Strand, London WC2R 2LS, UK}
\author{Malcolm Fairbairn}
\email{malcolm.fairbairn@kcl.ac.uk}\thanks{-- \href{https://orcid.org/0000-0002-0566-4127}{0000-0002-0566-4127}}
\affiliation{Physics, King's College London, Strand, London WC2R 2LS, UK}
\author{Diego Blas}
\email{diego.blas@gmail.com}\thanks{-- \href{https://orcid.org/0000-0003-2646-0112}{0000-0003-2646-0112}}
\affiliation{Grup de F\'{i}sica Te\`{o}rica, Departament de F\'{i}sica, Universitat Aut\`{o}noma de Barcelona, 08193 Bellaterra, Spain}
\affiliation{Institut de Fisica d’Altes Energies (IFAE), The Barcelona Institute of Science and Technology, Campus UAB, 08193 Bellaterra (Barcelona), Spain}
\author{Xiaolong Du}
\email{xdu@carnegiescience.edu}\thanks{-- \href{https://orcid.org/0000-0003-0728-2533}{0000-0003-0728-2533}}
\affiliation{Carnegie Observatories, 813 Santa Barbara Street, Pasadena, CA 91101, USA}
\affiliation{Department of Physics and Astronomy, University of California, Los Angeles, CA 90095, USA}
\author{David J. E. Marsh}
\email{david.j.marsh@kcl.ac.uk}\thanks{-- \href{https://orcid.org/0000-0002-4690-3016}{0000-0002-4690-3016}}
\affiliation{Physics, King's College London, Strand, London WC2R 2LS, UK}

\begin{abstract} 

If dark matter is composed of axions, then axion stars form in the cores of dark matter halos.  These stars are unstable above a critical mass, decaying to radio photons that heat the intergalactic medium, offering a new channel for axion indirect detection. We recently provided the first accurate calculation of the axion decay rate due to axion star mergers. In this work we show how existing data concerning the CMB optical depth leads to strong constraints on the axion photon coupling in the mass range $10^{-14}\,{\rm eV}\lesssim m_a\lesssim 10^{-8}\,{\rm eV}$. Axion star decays lead to efficient reionization of the intergalactic medium during the dark ages. By comparing this non-standard reionization with Planck legacy measurements of the Thomson optical width, we show that couplings in the range $10^{-14}\,{\rm GeV}^{-1} \lesssim g_{a\gamma\gamma} \lesssim 10^{-10}\,{\rm GeV}^{-1}$ are excluded for our benchmark model of axion star abundance. Future measurements of the 21cm emission of neutral hydrogen at high redshift could improve this limit by an order of magnitude or more, providing complementary indirect constraints on axion dark matter in parameter space also targeted by direct detection haloscopes.

\end{abstract}

\preprint{KCL-PH-TH-2023-16, CERN-TH-2023-029}

\maketitle

{
  \hypersetup{linkcolor=black}
}

\section{Introduction}

\begin{figure}[t]
\centering
\hspace{-0.3cm} \includegraphics[width=0.48\textwidth]{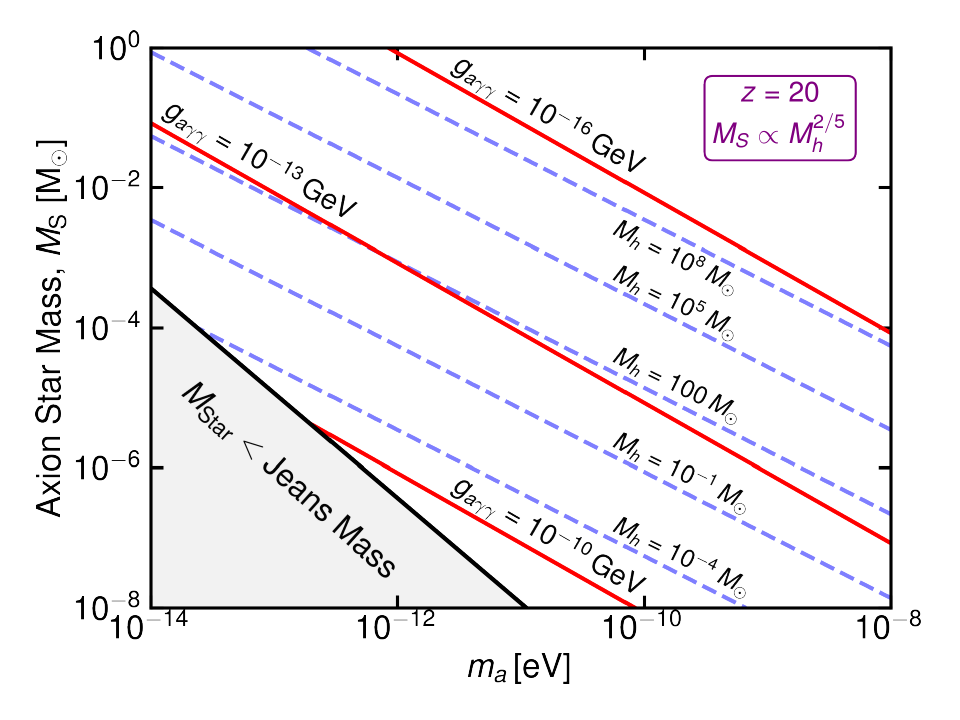}
\vspace{-0.7cm}
\caption{In blue dashed: isocontours of halo masses that host a given axion star mass at redshift $z = 20$ depending upon the value of the axion mass for the benchmark core halo mass relation $M_S\propto M_h^{2/5}$. In red we show isocontour lines of critical axion stars ($M_S^{\rm decay}$) as a function of several $g_{a\gamma\gamma}$ values.}\label{fig:MCrit_Core-Halo} 
\end{figure}

Numerical simulations of axion-like dark matter cosmologies have shown that a solitonic core forms in the center of every dark matter halo, see~\cite{Schive:2014dra,Schive:2014hza} for the first simulations and~\cite{Schwabe:2016rze,Mocz:2017wlg,Veltmaat:2018dfz,Nori:2020jzx,Mina:2020eik,Chan:2021bja} for further studies. 
In Ref.~\cite{Du:2023jxh} we have explicitly calculated the number and energy density evolution of axion stars from hierarchical structure formation using semi-analytic models starting from an initial adiabatic perturbation spectrum. Ref.~\cite{Schive:2014hza} demonstrated a power law relation between the mass of an axion star and its host halo:
\begin{align}
M_{S} = \left[\frac{M_{h}}{M_{\rm min}(z)}\right]^{\alpha}M_{\rm min}(z)\,,
\label{eq:Core-Halo_Relation}
\end{align}
where $z$ is the redshift, $M_h$ is the virial halo mass, $\alpha$ is a power law exponent, and~\cite{Schive:2014hza}:
\begin{align}
M_{\rm min}(z)=\,&1.4\times10^{-6}\left(\frac{m_a}{10^{-13}\text{ eV}}\right)^{-3/2} \nonumber \\
                &\left[\zeta(z)/\zeta(0)\right]^{1/4}(1+z)^{3/4}M_{\odot}\,,
\label{eqn:Mmin}
\end{align}
where $m_a$ is the axion-like dark matter mass and  $M_{\rm min}$ is the smallest halo mass that could host a soliton of mass $M_S$ at a given redshift (given by the Jeans scale)~\cite{Khlopov:1985jw,Hu:2000ke}. Ref.~\cite{Schive:2014hza} found a core-halo mass relation of the form $M_S\propto M_h^{1/3}$ while Refs.~\cite{Mocz:2017wlg,Nori:2020jzx,Mina:2020eik} have found $M_S \propto M_h^{3/5}$. Refs.~\cite{Chan:2021bja,Zagorac:2022xic} suggest that there is an intrinsic diversity in the core-halo mass relation. Importantly, all numerical simulations of axion-like dark matter cosmologies show that the core-halo mass relation is bounded between $\alpha = 1/3$ and $\alpha = 3/5$. The Monte Carlo merger tree model we employed in Ref.~\cite{Du:2023jxh} shows that an average slope $\alpha = 2/5$ captures this diversity well for both the soliton mass function and, more importantly, the merger rate. In Figure~\ref{fig:MCrit_Core-Halo} we show isocontours of constant $M_h$ for various values of $M_S$ and $m_a$ for the case $\alpha = 2/5$. For the other cases we refer the reader to Appendix~\ref{sec:othercorehalomass}. {We note that while the simulations of Refs.~\cite{Schive:2014dra,Schive:2014hza,Schwabe:2016rze,Mocz:2017wlg,Veltmaat:2018dfz,Nori:2020jzx,Mina:2020eik,Chan:2021bja} are performed for $m_{a}\sim 10^{-22}\,{\rm eV}$ their results should apply to any axion mass for three reasons: 1) the axion stars appear to form during the gravitational timescale of the halo, 2) the primordial power spectrum expected from inflation is almost scale-invariant, and 3) the evolution of non-relativistic axion dark matter features a scaling symmetry that allows extrapolation to any axion mass, see e.g.~\cite{Schive:2014hza}.}
\begin{figure*}[t]
\centering
\includegraphics[width=0.95\textwidth]{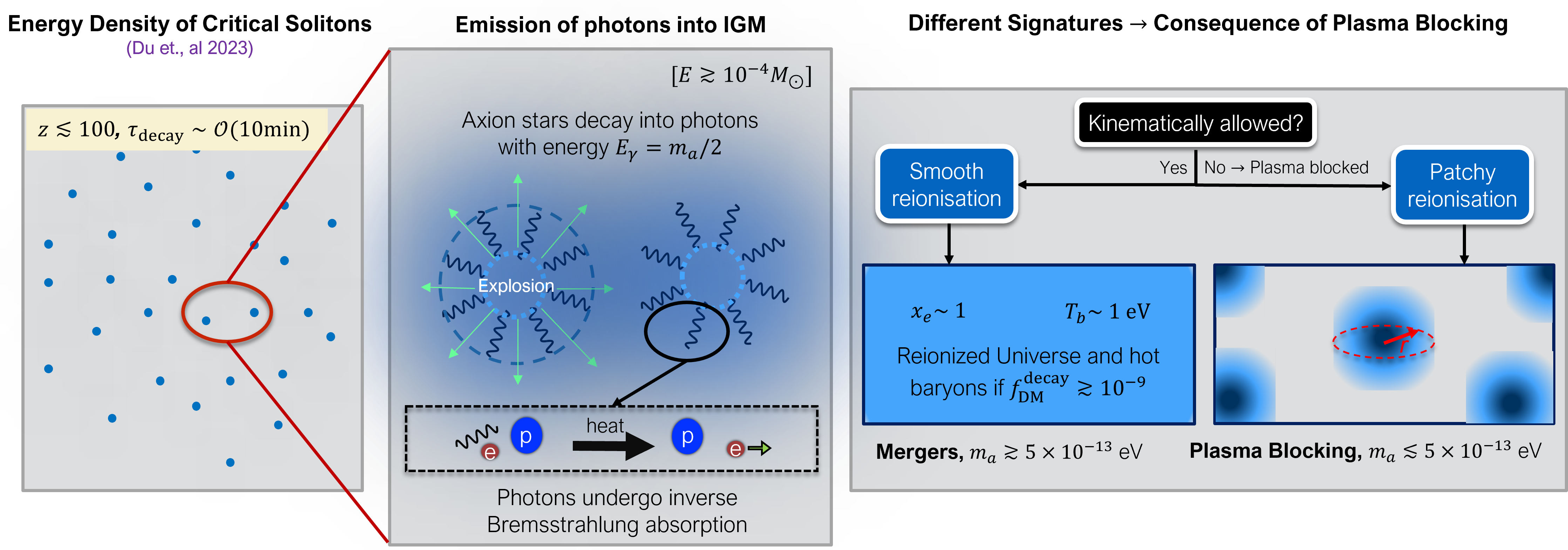}
\vspace{0.4cm}
\caption{Schematic of reionization caused by axion star explosions. Appreciable formation of dark matter halos hosting sufficiently massive axion stars occurs for $z\lesssim 100$, see~\cite{Du:2023jxh}. When an axion star decays, it releases a huge number of low-energy photons, which are absorbed by inverse Bremsstrahlung, leading to heating of the IGM. If the IGM becomes hot enough, reionization occurs. Energetically, reionization requires a fraction of around $f_{\rm DM}^{\rm decay}\approx 10^{-9}$ of the dark matter to decay. For the lowest mass axion-like particles, $m_a\lesssim 5\times 10^{-13}\text{eV}$, axion star decay is kinematically blocked at early times, leading to a population of super critical stars which decay all at once in a burst once the plasma frequency falls low enough to allow the decay. This leads to patchy reionization. At higher axion masses, plasma blocking is not efficient at the relevant redshifts, and instead super-critical stars formed by major mergers decay immediately as they form, leading to a more uniform and continuous reionization history. \label{fig:cartoon}}
\end{figure*}

\section{Axion Star Explosions}

Solitonic cores, also referred to as \emph{axion stars} have ultrahigh phase-space occupation numbers that can trigger collective processes which cannot occur in vacuum. Parametric resonance can lead to exponentially fast decays of axion stars into photons~\cite{Levkov:2020txo,Hertzberg:2018zte,Carenza:2019vzg,Amin:2020vja,Arza:2020eik,Hertzberg:2020dbk,Chung-Jukko:2023cow} or into relativistic axions~\cite{Eby:2015hyx,Eby:2016cnq,Levkov:2016rkk}, meaning axion stars become unstable above a certain mass~\cite{Levkov:2020txo,Chung-Jukko:2023cow} (vector dark matter solitons have a similar instability~\cite{Amin:2023imi}):
\begin{equation}\label{eq:Mdecay}
M_{S}^{\rm decay}\simeq 8.4\times 10^{-5}\,M_\odot\left(\frac{10^{-11}\,\text{GeV}^{-1}}{g_{a\gamma\gamma}}\right) \left( \frac{10^{-13}\,\text{eV}}{m_a}\right) \,,
\end{equation}
where $g_{a\gamma\gamma}$ is the axion-photon coupling. Super-critical solitons explode  into photons with $E_\gamma = m_a/2$. This happens by a collective process where photons produced by axion decays stimulate other axions to decay such that the axion star decay happens on a short timescale given by the light crossing time of the soliton: 
\begin{align}\label{eq:taudecay}
\!\!\tau_{S}^{{\rm decay}} \simeq r_c \simeq {\rm day} \left(\frac{8.4 \times 10^{-5} M_\odot}{M_S} \right) \left(\frac{10^{-13}\,{\rm eV}}{m_a}\right)^2\,.
\end{align}
The energy released in the explosion of a supercritical soliton is $E = M_S-M_S^{\rm decay}$. There are, however, two factors that could prevent this explosion:

1) If the axion-like particle possesses self interactions, then axion stars can decay into axions leading a ``Bosenova'' above a critical mass $M_{\rm B-Nova}$ set by the quartic self coupling $\lambda$~\cite{Eby:2015hyx,Eby:2016cnq,Levkov:2016rkk,Helfer:2016ljl,Chavanis:2011zm,Chavanis:2022fvh}. For a quartic coupling typical of a cosine instanton potential, $\lambda = m_a^2/f_a^2$, and axion-photon coupling $g_{a\gamma\gamma} \simeq \alpha_{\rm EM}/(2\pi f_a)$, one finds that $M_{S}^{\rm decay} \simeq 600 M_{\rm B-Nova}$. This means that for nominal values of $g_{a\gamma\gamma}$ the axion stars will not actually decay into photons but rather into relativistic axions, with possible limits from such a decay explored in Ref.~\cite{Fox:2023xgx}, see also~\cite{Eby:2021ece}. However, there are many models that feature significantly enhanced $g_{a\gamma\gamma}$ interactions, see e.g.~\cite{Farina:2016tgd,Agrawal:2017cmd,DiLuzio:2017pfr,Daido:2018dmu,Choi:2020rgn,Plakkot:2021xyx,Sokolov:2022fvs}, or suppressed quartic interactions, see e.g.~\cite{DiLuzio:2021pxd,DiLuzio:2021gos}. In what follows, we will consider that the relevant effect is decay into photons and thus our results will apply to scenarios such as those previously mentioned. 

2)  In the parameter space of interest for our study, axion stars are always hosted by halos of mass $M_h < 10^5\,M_\odot $ which means that they lie below the baryonic Jeans scale and are too light to capture any significant amount of baryons from the intergalactic medium (IGM)~\cite{Tegmark:1996yt,OLeary:2012gem}. This means that on average the axion stars formed will simply be surrounded by a baryon environment with average IGM properties. Nevertheless, the IGM contains ionized particles, which could kinematically block the axion decay by plasma effects. 
More concretely, the photon gets an effective mass-squared that is proportional to the number of free electrons in the plasma~\cite{Raffelt:1987im}:
\begin{align}\label{eq:omega_p}
    \omega_p^2(z) = \frac{4\pi\alpha_{\rm EM} \, n_e(z)}{m_e}\,.
\end{align}
Axion star decay is blocked until the plasma frequency drops to $\omega_p(z_{\rm decay}) < m_a/2$, resulting in an explosion of \emph{all} the axion stars with $M_S > M_S^{\rm decay}$ at specific redshift (see Appendix~\ref{sec:photon_plasma}): 
\begin{align}\label{eq:z_decay}
z_{\rm decay}  \simeq 32 \,\left(\frac{m_a}{10^{-13}\,{\rm eV}}\right)^{2/3}-1\,.
\end{align}

Once the plasma frequency has dropped such that parametric resonance decay is no longer blocked, then supercritical axion stars will explode as soon as they are formed. In Ref.~\cite{Du:2023jxh} we calculated the cosmological evolution of the number/energy density of critical axion stars using the extended Press-Schechter formalism and Monte Carlo merger trees. We considered the case that an axion star will only explode either if it is plasma blocked until it is super critical, or if it is formed by a \emph{major merger} defined by the merger of two axion stars of comparable mass, such that mass increase happens rapidly. 

Soliton explosions lead to axion dark matter decay, and can inject energy into the IGM comparable to or greater than energy emission from astrophysical processes, such as core collapse supernovae~\cite{Du:2023jxh}, and thus offer a new method of indirect detection of axions.

\section{Heating the Intergalactic Medium}

Figure~\ref{fig:cartoon} outlines qualitatively the mechanism just described: axion stars start to form and grow when halos form and merge, which happens appreciably only at $z\lesssim 100$. Once an axion star becomes massive enough and the plasma frequency is low enough, parametric resonance can take place and the star rapidly explodes into low energy photons of $E_\gamma = m_a/2$. These photons are absorbed by the IGM, heat it, and in turn via collisional ionizations lead to an increased ionization fraction of the Universe. 

Axion star decay results in the release of a huge number of low energy photons with $E_\gamma =m_a/2$. We will be interested in $m_a < 10^{-8}\,{\rm eV}$, whose associated low energy photons are efficiently absorbed via inverse Bremsstrahlung on ionized particles in the intergalactic medium, namely via $\gamma \, e^-\, p \to e^- p $ processes~\cite{Chluba:2015hma}. The net result of this absorption is to heat the IGM. In turn, once the temperature of the IGM goes above $T\sim 1\,{\rm eV}$, collisional ionization processes $e^- \, H \to 2e^- \, p $ ionize the Universe. As a result, the decay of axion stars may result into a period of early reionization, which is strongly constrained by Planck legacy data~\cite{Planck:2016mks,aghanim2020planck}. This forms the basis of the constraints derived in this paper, see also~\cite{McDermott:2019lch,Witte:2020rvb,Caputo:2020bdy,Bolliet:2020ofj,Slatyer:2015jla,Liu:2020wqz,Capozzi:2023xie,Liu:2023fgu} for constraints of this type for other dark matter candidates. 

Importantly, depending upon the absorption length scale of the photons produced by axion stars the effect on the IGM can be different. Considering the $\gamma \, e^-\, p \to e^- p $ absorption length, see~\cite{Chluba:2015hma} and Appendix~\ref{sec:absorption}, we find that for $m_a \lesssim 5\times 10^{-13}\,{\rm eV}$ the photons are absorbed within very small volumes and this generates a shockwave very much like the one formed by supernova explosions as the amount of energy released is very similar~\cite{Ostriker:1988zz}. On the other hand, for $m_a \gtrsim 5\times 10^{-13}\,{\rm eV}$ the absorption length-scale is larger than the inter-separation of axion stars results in a homogeneous cosmological increase in the global IGM temperature. It is important to highlight that in principle a fraction as small as $f_{\rm DM} \sim 3\times 10^{-9} \simeq 13.6\,{\rm eV}\, n_b/\rho_{\rm DM}$ of dark matter converted into heat in the IGM would be enough to fully reionize the Universe, see Appendix~\ref{sec:dfdzstuff}.

\section{Methodology}

We model the IGM temperature by including the heating generated by axion star decays and accounting for cooling from Compton processes, { collisional excitations and ionizations, and the expansion of the Universe}. {Collisional excitations and ionizations dominate in our scenario which we implement using the fitted rates in~\cite{Bolton:2006pc,doi:10.1063/1.555700}}. We then track the free electron density $x_e$ using the effective 3-level atom approximation~\cite{Peebles:1968ja, Zeldovich:1968qba}, including free protons and ${\rm He}^+$. We use the same rates and expressions as the recombination code RECFAST including their fudge factor~\cite{Seager:1999km, Seager:1999bc} --  modulo that we change $T_{b} \to T_{\gamma}$ in the terms in Eqs. (1) and (2) of~\cite{Seager:1999bc} that account for photon ionization as the approximation $ T_b \simeq  T_\gamma$ breaks down for $z < 200$. Finally, in the regime $m_a \lesssim 5\times 10^{-13}\,{\rm eV}$ the photons from axion star explosions are absorbed on scales smaller then their typical hosting halo and generate shockwaves which ionize small patches of the Universe. We track such explosions until they are unable to ionize more IGM. We follow standard methods used for supernova remnants~\cite{Ostriker:1988zz,1993ApJ...417...54T,Pieri:2006bd,Blas:2020nbs}, and find results very close to those expected from the Sedov-Taylor solutions~\cite{taylor,sedov}. We refer the reader to Appendices~\ref{sec:xe_Tb_evolution} and~\ref{sec:evolution_patchy} where we describe in detail all the relevant heating and ionization for the two regimes of interest, respectively.

\section{Constraints from the CMB Optical Depth}

\begin{figure}[t]
\centering
\hspace{-0.3cm} \includegraphics[width=0.49\textwidth]{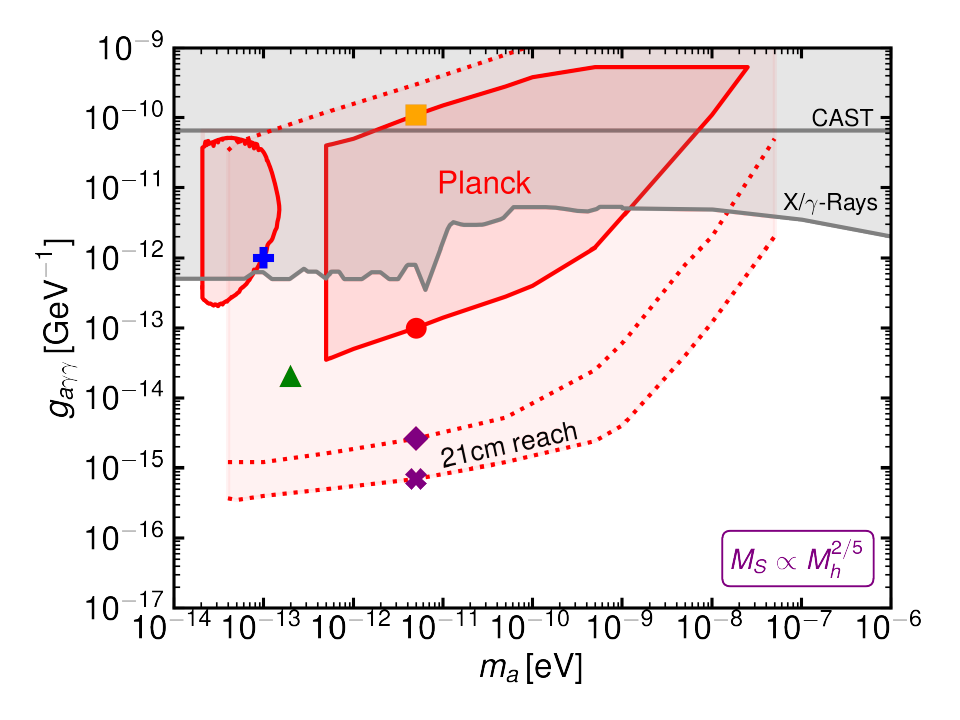}
\vspace{-0.9cm}
\caption{Parameter space excluded by Planck measurements of the Thomson optical depth. We highlight also reach of future 21cm surveys and constraints from X/$\gamma$-Ray observations~\cite{Reynolds:2019uqt,Dessert:2020lil,Meyer:2020vzy,Reynes:2021bpe,Dessert:2022yqq,Noordhuis:2022ljw} and CAST~\cite{CAST:2017uph}, see~\cite{Oharerep}. The evolution of the free electron fraction and the baryon temperature is shown for the points highlighted by colored symbols in Figure~\ref{fig:xe_Tb_cases}.} \label{fig:constraints}
\end{figure}

\begin{figure}[t]
\centering
\hspace{-0.3cm} \includegraphics[width=0.49\textwidth]{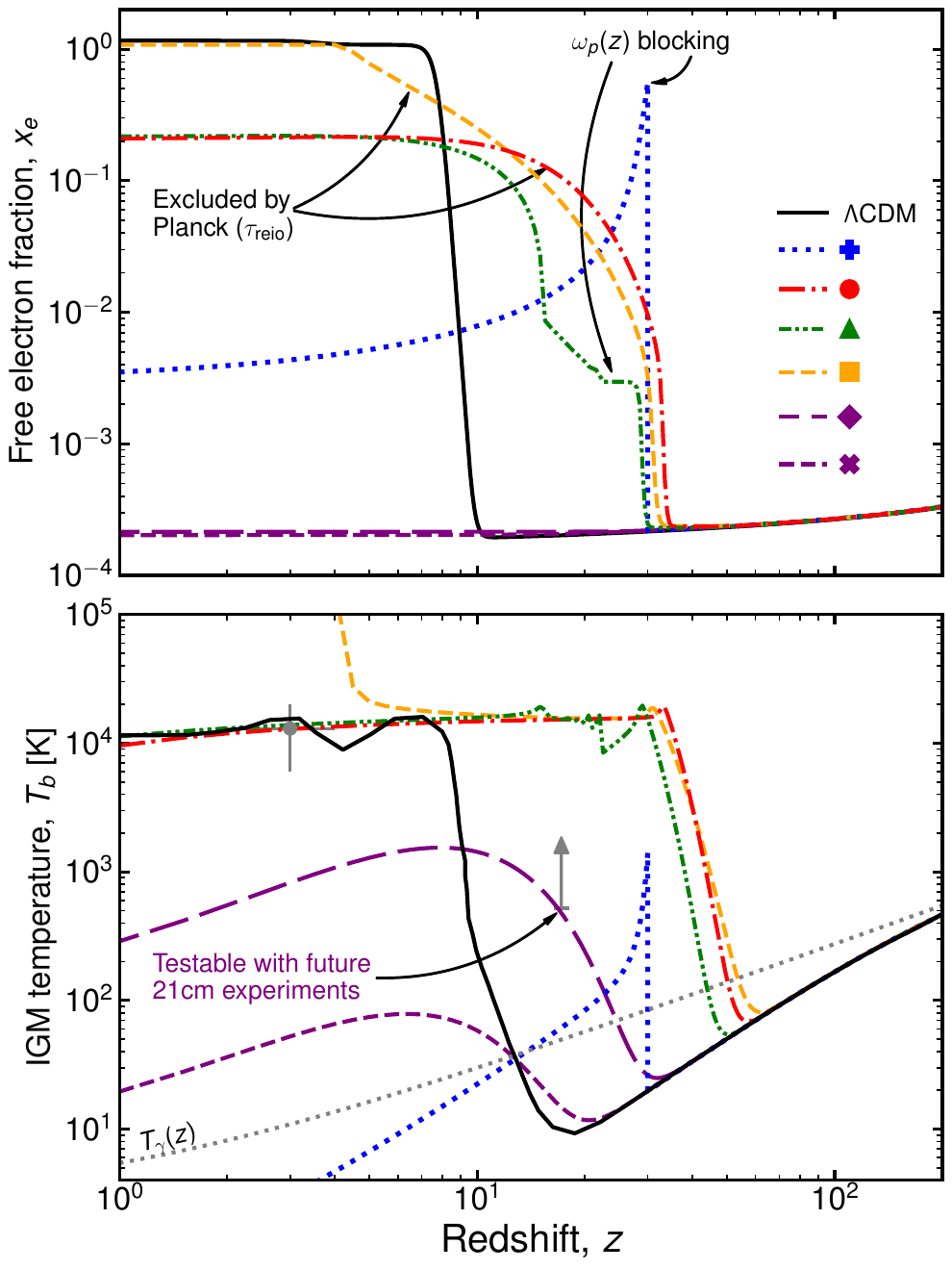}
\vspace{-0.6cm} 
\caption{Evolution of $x_e$ (upper) and $T_b$ (lower) for the six scenarios highlighted in Figure~\ref{fig:constraints} with symbols. For the scenarios we show there are no astrophysical sources of heating or reionization. We also show in black the $\Lambda$CDM evolution for $x_e$ from the Planck best fit cosmology and $T_b$ from the fiducial model of~\cite{McQuinn:2015icp}. In the lower panel we highlight scenarios that can be tested with future 21cm observations, conservatively (long dashed) and optimistically (dashed).}\label{fig:xe_Tb_cases}
\end{figure}

 We derive conservative constraints on reionization driven by axion star explosions by using the $95\%$ upper limit on the integrated Thomson optical depth to reionization from Planck CMB observations: $\tau_{\rm reio}< 0.068$~\cite{Planck:2016mks,aghanim2020planck}. In practice, and as customarily done for other types of energy injections~\cite{Liu:2020wqz,Capozzi:2023xie,Liu:2023fgu}, since observations show that the Universe should be fully ionized by redshift $z\simeq 6$~\cite{Fan:2005es}, and this implies a minimum contribution of $\tau_{\rm reio} = 0.0384$, we then consider a region of parameter space excluded if the contribution to the optical width to reionization from axion star explosions at $50>z>6$ is $\tau_{\rm reio}> 0.03$. We note that this is a conservative approach as it assumes the Universe was not reionized until $z = 6$ by standard astrophysical sources. 

The axion parameter space excluded as a result of this constraint is shown in the two regions in dark red in Figure~\ref{fig:constraints}, and the resulting evolution of the ionization fraction and IGM temperature for representative models is shown in Figure~\ref{fig:xe_Tb_cases}. Constraints in the region $m_a\lesssim 5\times 10^{-13}\,{\rm eV}$ are a result of patchy reionization generated by the decay of all critical axion stars in the Universe once the axion decay is kinematically allowed, namely when $\omega_p(z_{\rm decay}) = m_a/2$. For $m_a\gtrsim 5\times  10^{-13}\,{\rm eV}$ the photon plasma mass is small compared to the axion mass, and the constraints arise instead from super-critical axion stars formed via mergers providing continuous heating and uniform reionization. The intermediate region of parameter space is constrained only at $1\sigma$ because in this regime the plasma frequency rises and blocks further axion star decay, see green curve of the upper panel of Figure~\ref{fig:xe_Tb_cases}. This region of parameter space is, however, expected to be tested with large scale CMB polarization data from LiteBIRD~\cite{LiteBIRD:2020khw}.

We note that the Planck constraints do not extend to arbitrarily large $g_{a\gamma\gamma}$ - the larger the value of $g_{a\gamma\gamma}$, the smaller the mass of the halos that host critical soliton stars. In axion-like dark matter cosmologies structure formation is suppressed at $M_h< M_{\rm min}$, see Eq.~\eqref{eqn:Mmin}, and couplings $g_{a\gamma\gamma }\gtrsim 10^{-10}\,{\rm GeV}$ correspond to axion stars hosted in halos whose abundance is suppressed (which may in turn lead to further cosmological constraints, see e.g.~\cite{Irsic:2019iff,Rogers:2020ltq}). In addition, we notice that the bounds change their shape for $m_a \gtrsim 5\times 10^{-11}\,{\rm eV}$. For this region of parameter space the photon absorption efficiency in the IGM is smaller than one and this weakens the constraints.  

\vspace{-0.4cm}

\section{Lyman-$\alpha$ and Spectral Distortions} 

We considered two other existing measurements that can limit this scenario for IGM heating. Since axion star explosions occur when Compton scattering is inefficient ($z\lesssim 10^4$), they generate $y$-type distortions of the CMB energy spectrum. The change to the photon energy density from such an effect must satisfy $\delta \rho_\gamma/\rho_\gamma \lesssim 5\times 10^{-5}$ based on COBE/FIRAS measurements~\cite{Fixsen:1996nj,Bolliet:2020ofj}, which translates to $f_{\rm DM}^{\rm decay} \lesssim 2\times 10^{-7}$ for decaying DM: significantly weaker than the Planck optical depth constraint. At redshifts $ 2 \lesssim z \lesssim 7$ the Lyman-$\alpha$ forest can measure the IGM temperature see e.g.~\cite{Walther:2018pnn,Gaikwad:2020art,Muller:2020pib}. The IGM temperature at these redshifts satisfies $T_b \lesssim 1.5\times 10^{4}\,{\rm K}$ which can also be used to constrain axion star decays. This is shown as the upper contour in Figure~\ref{fig:constraints}, with an accompanying $T_b(z)$ in Figure~\ref{fig:xe_Tb_cases} (bottom panel). Lyman-$\alpha$ data could be competitive with the Planck optical depth constraint, although a dedicated study including other sources of heat and cooling mechanisms would be needed to set definitive constraints.

\section{21 cm Cosmology} 

 Measurements of the hyperfine 21cm transition are expected to revolutionize our understanding of the cosmic dawn and the epoch of reionization, see~\cite{Pritchard:2011xb,Furlanetto:2006jb} for reviews. There have already been several interesting bounds on the emission power of the 21cm line at various redshifts~\cite{HERA:2022wmy,Singh:2021mxo}, and even a putative detection~\cite{Bowman:2018yin} (which is disputed see, e.g.~\cite{Singh:2021mxo}). The observational status of the cosmological 21cm line is still in its infancy but it is expected that ongoing and upcoming experiments such as HERA~\cite{DeBoer:2016tnn} and SKA~\cite{Mellema:2012ht} among others should be able to robustly detect the 21cm line and provide a view of the thermal state of the Universe at redshifts $4\lesssim z\lesssim 30$~\cite{Barry:2021szi}, see~\cite{Villanueva-Domingo:2021vbi} for a compilation of experiments targeting this line. 

As highlighted in Figure~\ref{fig:xe_Tb_cases}, axion star explosions greatly enhance the IGM temperature at the redshifts where the 21cm line will be targeted $z\lesssim 35$. In particular, the rise of $T_b$ at large redshifts $z\gtrsim 20$ is a feature that is not easy to achieve astrophysically, see e.g.~\cite{Cohen:2016jbh}. The sky-averaged 21 cm brightness temperature can be written as~\cite{Pritchard:2008da}:
\begin{align}
    \!\!\! T_{21} = 27\,{\rm mK} \,x_{\rm HI}\, \frac{\Omega_b h^2}{0.023}\left[\frac{0.15}{\Omega_m h^2}\frac{1+z}{10}\right]^{1/2}\left[1-\frac{T_\gamma}{T_S} \right]\,,
\end{align}
where $x_{\rm HI}$ is the fraction of neutral hydrogen in the Universe and $T_S$ is the spin temperature. At $z \lesssim 25$ the spin temperature is $T_S\simeq T_b$~\cite{Mesinger:2010ne} and we then see that for $T_{b} > T_\gamma$ the 21cm signal could reach a maximum in emission of $\sim 35\,{\rm mK} $ at $z \simeq 20$ provided $T_b \gg T_\gamma$. This is in strong contrast with what is expected in most cold IGM models where $T_{b}(z=20) \sim 7\,{\rm K}$ and which would lead to a strong absorption feature with $T_{21}\sim -200\,{\rm mK}$. The sensitivity of SKA in this redshift range is expected to be $\Delta T_{21} \sim 10\,{\rm mK}$, which means that SKA should be able to clearly differentiate between these two cases and thus future 21cm observations will test scenarios where axion star explosions heat the IGM. 

In Figure~\ref{fig:constraints} we highlight in light red the region of parameter space that could be tested by an experiment such as SKA by demanding that $T_{21} < 30\,{\rm mK}$ at $z \simeq 20$ (equivalent to $T_b \gtrsim 500\,{\rm K}$ by $z \simeq 20$). A tighter limit can be arrived at if one could differentiate any model from $\Lambda$CDM that causes $T_b>T_\gamma$ at $z\gtrsim 10$. This is shown by the bottom light line in Figure~\ref{fig:constraints}. 

\section{Conclusions} 

 Parametric-resonance instability of axion stars leads to an enhanced decay rate of axion dark matter into low energy radio photons. We have shown how the production of such photons heats the IGM, leads to reionization, and alters the optical depth of CMB photons. Planck legacy measurements of the optical depth form the basis of our new constraints on $g_{a\gamma\gamma}$ highlighted in Figure~\ref{fig:constraints}, which is the strongest limit on axion dark matter in the relevant mass range by more than an order of magnitude. Future 21cm measurements of the IGM during the dark ages could improve this limit by more than an order of magnitude. The overlap of these indirect probes with the target regions of the DM-Radio haloscope program~\cite{DMRadio:2022pkf} invites further careful consideration of axion star explosions as a new tool in the hunt for dark matter.

\section{Outlook} 

Aspects of this model that require further exploration include primarily a detailed study of soliton merger rates in simulations displaying a core-halo diversity in the redshift and particle mass ranges of interest. While our Monte Carlo merger trees in~\cite{Du:2023jxh} suggest that $M_S \propto M_h^{2/5}$ accounts well for diversity in the soliton merger rate only simulations in the redshifts of interests can fully support or modify this conclusion. {Furthermore, the cosmological simulations of axion stars are typically carried out for $m_a \sim 10^{-22}\,{\rm eV}$. In these simulations axion stars appear to form upon halo collapse, and given the almost scale-invariant primordial power spectrum and the fact that the Schrodinger-Poisson equations solved in them features a scaling symmetry, their results can then be extrapolated to any axion mass. Nevertheless, it would be very interesting to perform explicit simulations of axion star formation and growth within the mass range of interest for our study, $10^{-14}\,{\rm eV} \lesssim m_a \lesssim 10^{-6}\,{\rm eV}$.}

On the side of particle physics, the strongest assumption of our analysis is that either axion quartic couplings are suppressed or $g_{a\gamma\gamma}$ couplings are enhanced compared with canonical expectations. While models of these types exist, construction of further explicit axion-like dark matter models above the QCD line would be required to address how natural such couplings are. Nevertheless, simulations of axion star explosions including both quartic and photon couplings may yet find that Bosenova triggered photon parametric resonance can occur for more standard coupling ratios. 
Furthermore, determining the true reach of 21cm measurements requires a full simulation of the 21cm anisotropies including anomalous low-energy photon injection from axion stars. We suspect that anisotropies in this model will differ strongly from $\Lambda$CDM, and thus bounds may even be more powerful than our estimates.

%\begin{widetext}

\begin{acknowledgments}
 ME would like to thank Andrea Caputo, Josh Eby, Ivan Esteban, and Sam Witte for very useful discussions. MF gratefully received funding via the STFC particle theory grant STFC-ST/T000759/1.  CKP is funded via an STFC quota studentship.
DB is supported by a `Ayuda Beatriz Galindo Senior' from the Spanish `Ministerio de Universidades', grant BG20/00228. 
The research leading to these results has received funding from the Spanish Ministry of Science and Innovation (PID2020-115845GB-I00/AEI/10.13039/501100011033).
IFAE is partially funded by the CERCA program of the Generalitat de Catalunya. DJEM is supported by an Ernest Rutherford fellowship from STFC (NO. ST/T004037/1).
\end{acknowledgments}

\makeatletter

\appendix

\begin{figure*}[t]
\centering
\begin{tabular}{cc}
\includegraphics[width=0.48\textwidth]{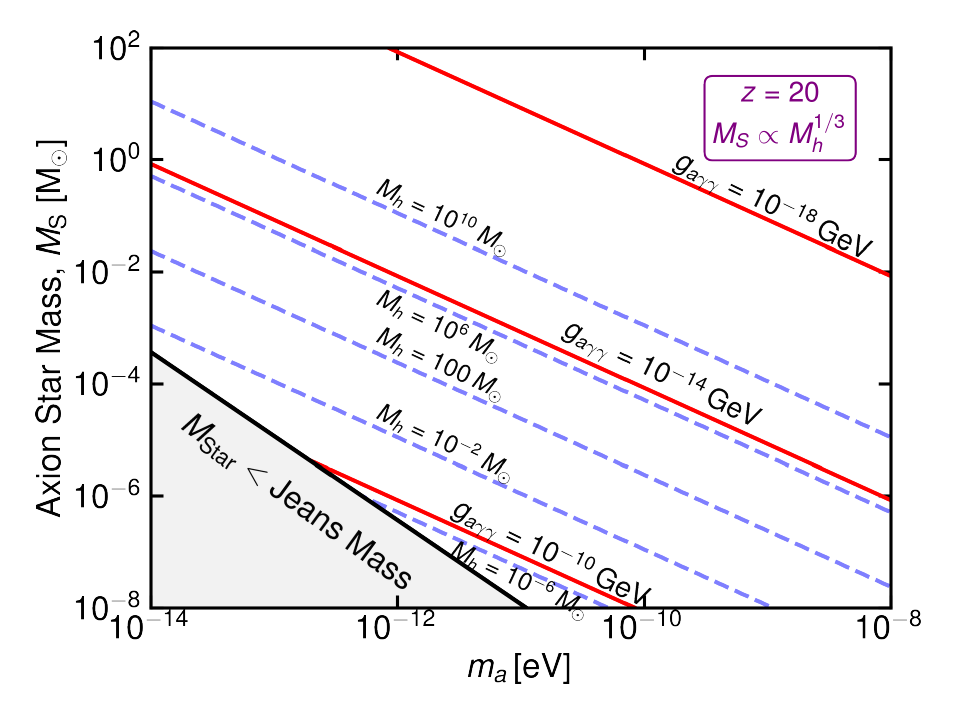} & \includegraphics[width=0.48\textwidth]{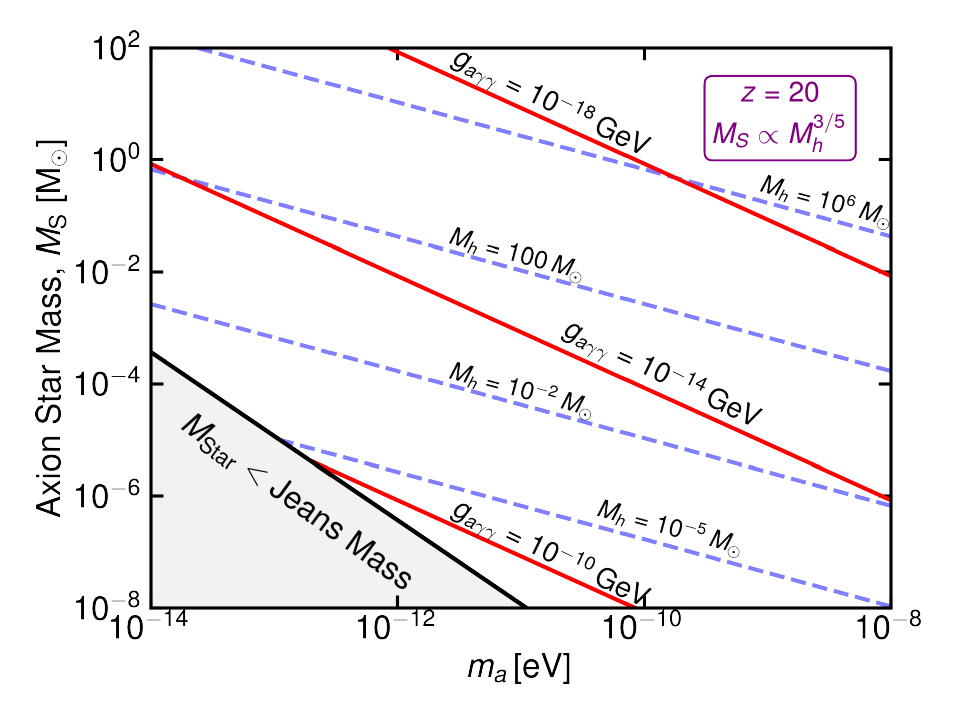}
\end{tabular}
\vspace{-0.3cm}
\caption{Isocontours of halo masses (dashed blue) that host an axion star mass (y axis) as a function of axion mass (x axis) at a redshift $z = 20$, see Eq.~\eqref{eq:Core-Halo_Relation}. In red we show the value of a critical axion star above which the star can explode into photons given an axion-photon coupling in red, see Eq.~\eqref{eq:Mdecay}. The left panel corresponds to the Schive relation, with $M_S \propto M_h^{1/3}$ while the right panel corresponds to the case $M_S \propto M_h^{3/5}$. We can clearly see that for a fixed axion star mass the mass of a halo hosting it is much smaller for the case $\alpha = 3/5$ than for the one with $\alpha = 1/3$. This in turn means that significantly smaller $g_{a\gamma\gamma}$ couplings can be probed. We highlight in grey regions of parameter space for which such axion stars cannot have possibly formed by $z = 20$ as their masses would lie below the effective Jeans mass generated by quantum pressure, see Eq.~\eqref{eqn:Mmin}.} \label{fig:Mcore-Mhalo_app}
\end{figure*}

%\end{widetext}

%\onecolumngrid

%\setcounter{figure}{0}
%\renewcommand{\thefigure}{S\arabic{figure}}

\begin{figure*}[t]
\centering
\begin{tabular}{cc}
\includegraphics[width=0.48\textwidth]{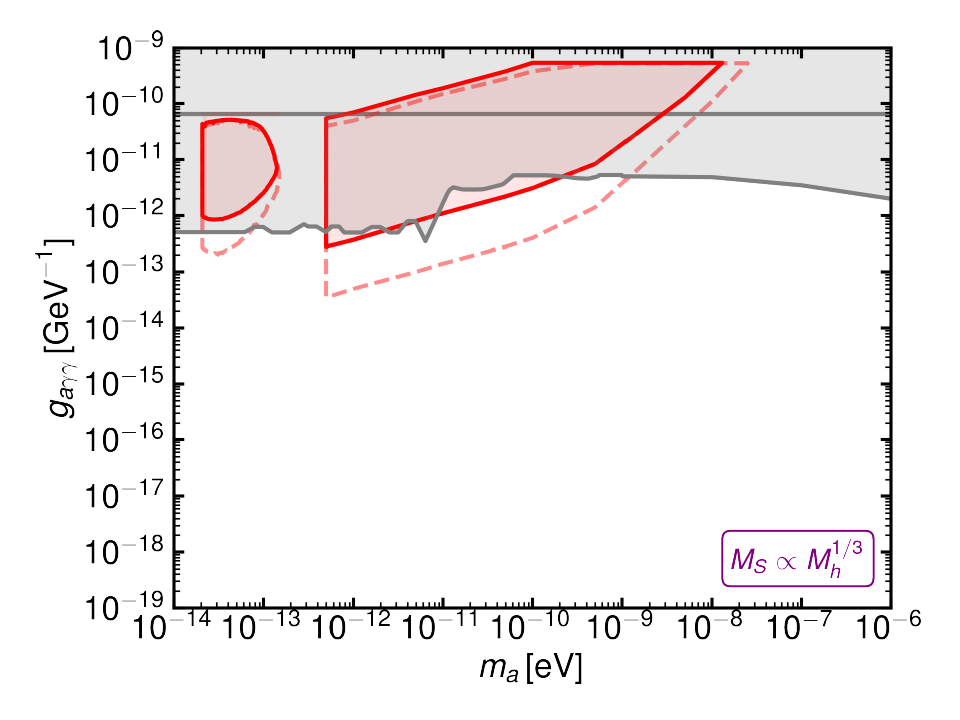} & \includegraphics[width=0.48\textwidth]{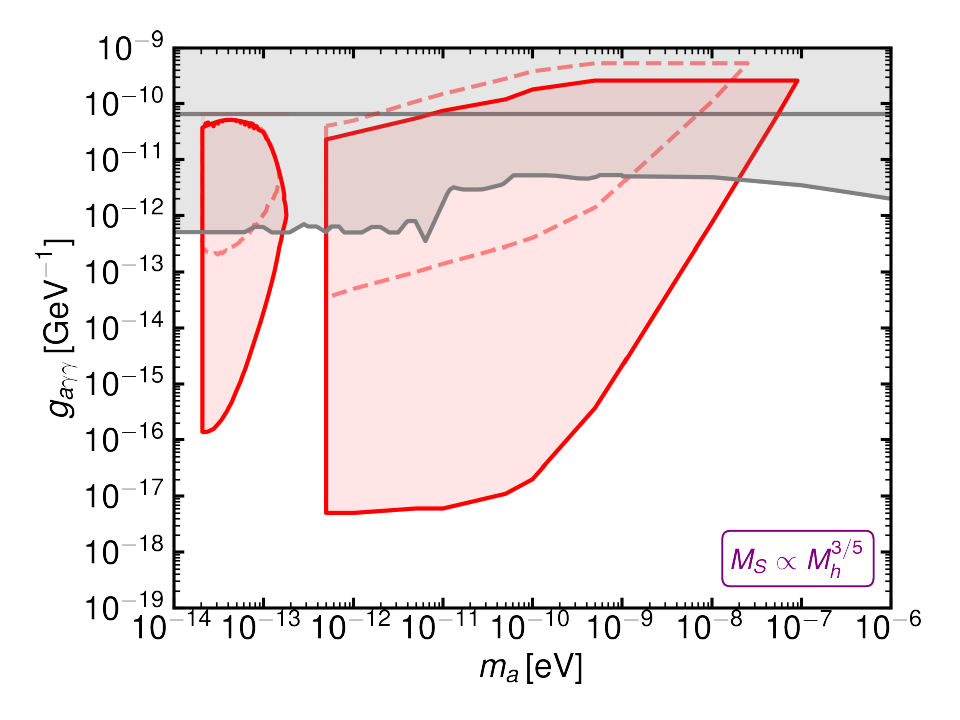}
\end{tabular}
\vspace{-0.3cm}
\caption{Equivalent to Figure~\ref{fig:constraints} but for two other core-halo mass relations $M_S \propto M_h^{1/3}$ (left) and $M_S \propto M_h^{3/5}$ (right). In light red dashed lines we show the benchmark case with $M_S\propto M_h^{2/5}$ for comparison.} \label{fig:constraints_13_35}
\end{figure*}

%%%%%%%%%%%%%%%%%%%%%%%%%%%%%%%%%%%%%%%%%%%%%%%%%%%%%%%%%%%%%%%%%%%%%%%%%%%%%%%%%%%%
\begin{center}
    {\bf Appendices} 
\end{center}
%%%%%%%%%%%%%%%%%%%%%%%%%%%%%%%%%%%%%%%%%%%%%%%%%%%%%%%%%%%%%%%%%%%%%%%%%%%%%%%%%%%%

\begin{itemize}[label={}]    
    \item  Appendix~\ref{sec:othercorehalomass}: Constraints for other core-halo mass relations in the parameter space $g_{a\gamma\gamma}$ and $m_a$.
    \item  Appendix~\ref{sec:dfdzstuff}: Decaying dark matter fraction from axion star explosions as a function of $g_{a\gamma\gamma}$ and $m_a$. 
    \item Appendix~\ref{sec:photon_plasma}: Evolution of the photon plasma mass and its effect on axion star decays. 
    \item Appendix~\ref{sec:absorption}: Absorption rate of photons generated from axion star explosions in the IGM. 
    \item Appendix~\ref{sec:xe_Tb_evolution}: Evolution equations for $x_e$ and $T_b$ in the presence of homogeneous and continuous energy releases from axion star mergers and explosions. Relevant for $m_a \gtrsim 5\times 10^{-13}\,{\rm eV}$.
    \item Appendix~\ref{sec:evolution_patchy}: Patchy reionization evolution generated by axion star explosions where $m_a \lesssim 5\times 10^{-13}\,{\rm eV}$. Evolution of shockwaves and ionized bubbles and its impact on the global $x_e$.
\end{itemize}

\vspace{-0.3cm}

%%%%%%%%%%%%%%%%%%%%%%%%%%%%%%%%%%%%%%%%%%%%%%%%%%%%%%%%%%%%%%%%%%%%%%%%%%
\section{Results for other Core-halo mass relations}\label{sec:othercorehalomass}
%%%%%%%%%%%%%%%%%%%%%%%%%%%%%%%%%%%%%%%%%%%%%%%%%%%%%%%%%%%%%%%%%%%%%%%%%%

\begin{figure*}[t]
\centering
\begin{tabular}{cc}
\includegraphics[width=0.48\textwidth]{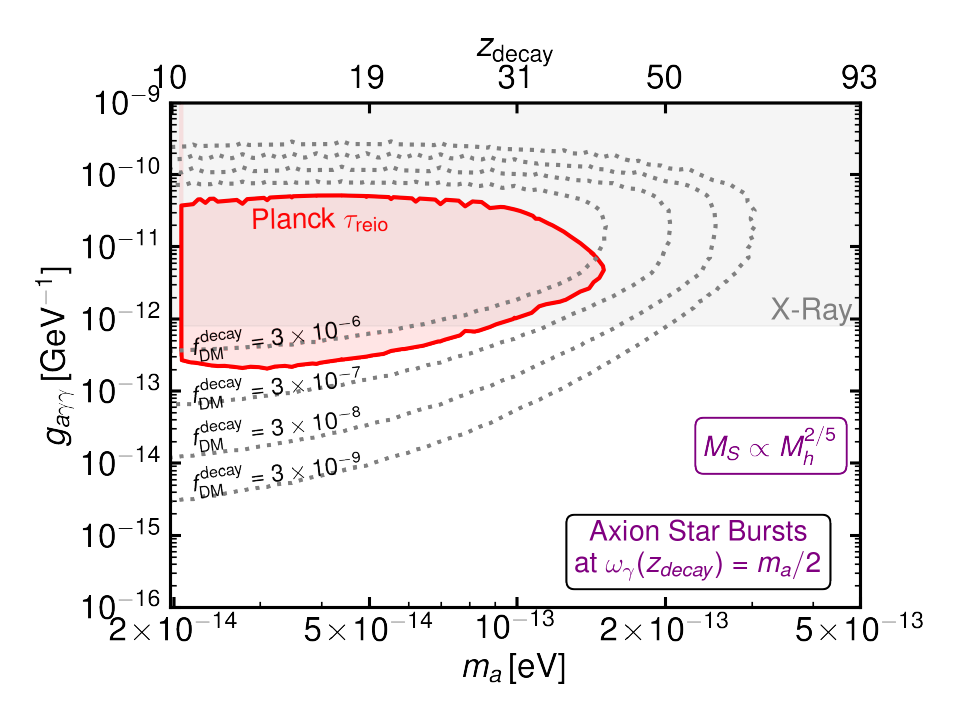} & \includegraphics[width=0.48\textwidth]{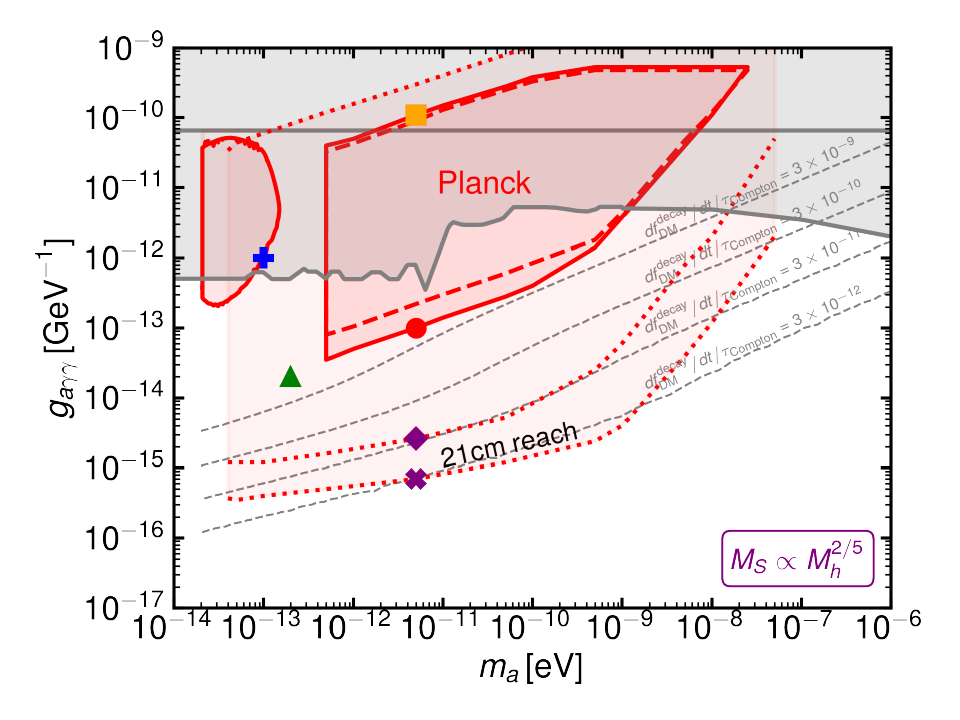}
\end{tabular}
\vspace{-0.3cm}
\caption{\textit{Left:} isocontours of $f_{\rm DM}^{\rm decay}$ compared with the constraints obtained from Planck. \textit{Right:} isocontours of $df_{\rm DM}^{\rm decay}/dt/\tau_{\rm Compton}$. We can clearly appreciate how the constraints follow the same shape and follow closely the case $3\times 10^{-9}$ as expected from energy arguments. The region $m_a\gtrsim 10^{-10}\,{\rm eV}$ deviates from this expectations because the probability of photon absorption in the IGM is smaller than 1. In addition, in dashed red we show regions of parameter space with $\tau_{\rm reio} = 0.13$, namely, with an optical depth that exceeds the Planck bound by a factor of 2.} \label{fig:constraints_app}
\end{figure*}

In the main text we have shown all results for the benchmark core-halo mass relation in Eq.~\eqref{eq:Core-Halo_Relation} with $\alpha = 2/5$. Here we show the results for other two benchmark core-halo mass relations: $\alpha = 1/3$ and $\alpha = 3/5$. The former was the first one found in the literature~\cite{Schive:2014dra,Schive:2014hza}. The case $\alpha = 3/5$ was found in Refs.~\cite{Mocz:2017wlg,Nori:2020jzx,Mina:2020eik}. In addition, Refs.~\cite{Chan:2021bja,Zagorac:2022xic} have shown evidence not for a strict core-halo mass relation but rather for diversity. Importantly, all numerical simulations show that axion stars lie somewhere in between $\alpha= 1/3$ and $\alpha = 3/5$. In addition, our Monte Carlo merger tree~\cite{Du:2023jxh} shows that $\alpha = 2/5$ captures well the effect of diversity in the merger rate of axion stars and in consequence represents our nominal choice for the core-halo mass relation. For completeness, in this appendix we show the resulting Planck constraints for these other two scenarios. In particular, Figure~\ref{fig:Mcore-Mhalo_app} shows the $M_S$-$M_h$ relation for the two cases while Figure~\ref{fig:constraints_13_35} shows the Planck CMB constraints for each of them.

%%%%%%%%%%%%%%%%%%%%%%%%%%%%%%%%%%%%%%%%%%%%%%%%%%%%%%%%%%%%%%%%%%%%%%%%%%%%%%%%%%%%%%%%%%%%
\section{Decaying dark matter fraction and axion star merger evolution}\label{sec:dfdzstuff}
%%%%%%%%%%%%%%%%%%%%%%%%%%%%%%%%%%%%%%%%%%%%%%%%%%%%%%%%%%%%%%%%%%%%%%%%%%%%%%%%%%%%%%%%%%%%

Planck constraints on the axion parameter space are strongly related to the amount of decaying dark matter that is injected as heat in the IGM. There are two distinct regions of parameter space. For $m_a\lesssim 5\times 10^{-13}\,{\rm eV}$ there are three key effects: 1) axion stars cannot decay until the photon plasma mass is low enough, 2) the explosion thus occurs as a burst of energy when $\omega_p(z)<m_a/2$, and 3) the photons generated from axion star explosions are absorbed on very small volumes and this generates shockwaves and bubbles of ionized IGM. This leads to patchy ionization. In this case, the constraints on the parameter space closely resemble regions of constant decaying dark matter density evaluated at the redshift of decay, $z_{\rm decay}$. This is explicitly shown in the left panel of Figure~\ref{fig:constraints_app}. There we can see that the region of parameter space excluded by Planck closely follows a region of parameter space where $f_{\rm DM}^{\rm decay} \simeq 10^{-6}$. It is important to note that this is almost three orders of magnitude larger than what would energetically be needed if all the energy were to be injected homogeneously as heat in the IGM. However, since the observable $\tau$ is strongly sensitive to the volume of ionized IGM in the Universe this patchy reionization is much less efficient in fully ionizing the entire Universe. This is analogous to what happens in standard reionization where the most efficient sources of the global ionization of the Universe are UV and X-ray photons with mean free paths similar to the size of the observable Universe. 

In the region of parameter space with $m_a\gtrsim 5\times 10^{-13}\,{\rm eV}$ the absorption of photons from axion star explosions takes place over distances that are larger than the typical inter-separation of axion stars. That means that the process can effectively be seen as homogeneous. In addition, in this case the emission is continuous because the photon plasma mass is small compared to $m_a/2$. In this regime what matters is how much energy is released into heat and on which timescale does the IGM cool. The former we explicitly calculated in Ref.~\cite{Du:2023jxh}. The latter is simply given by Compton cooling and reads (see Eq.~\eqref{eq:Tm_ODE}):
\begin{align}
 \tau_{\rm Compton} = \frac{45}{4\pi^2} \frac{1}{T_\gamma^4} \frac{m_e}{\sigma_T} \simeq  \frac{2.3\times 10^{12}}{(1+z)^4} \,{\rm years}\,.
\end{align}
Thus, given energy arguments we expect regions of parameter space where $df_{\rm DM}^{\rm decay}/dt/\tau_{\rm Compton} \gtrsim 3\times 10^{-9}$ can lead to reionization of the Universe. This is explicitly shown in the right panel of Figure~\ref{fig:constraints_app} where we see that the Planck exclusion region is parallel and very close to this line. We notice that 21 cm observations will be sensitive to rates that are $\sim 2$ orders of magnitude smaller than those currently tested by Planck measurements of $\tau_{\rm reio}$.

%%%%%%%%%%%%%%%%%%%%%%%%%%%%%%%%%%%%%%%%%%%%%%%%%%%%%%%%%%%%%%%%%%%%%%%%%%%%%%%%%%%%
\section{Cosmological Photon Plasma Mass}\label{sec:photon_plasma}
%%%%%%%%%%%%%%%%%%%%%%%%%%%%%%%%%%%%%%%%%%%%%%%%%%%%%%%%%%%%%%%%%%%%%%%%%%%%%%%%%%%%

\begin{figure}[t]
\centering
\hspace{-0.3cm} \includegraphics[width=0.48\textwidth]{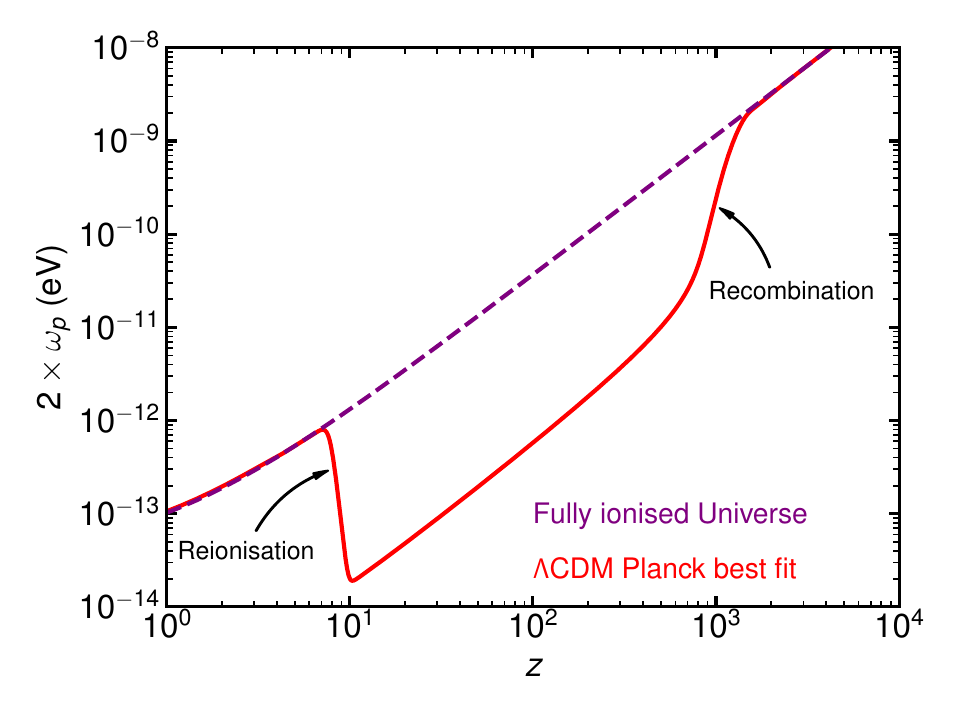}
\vspace{-0.6cm}
\caption{Minimum mass of an axion that can decay into photons as a function of redshift, $m_a > 2\, \omega_p(z)$. In red we show the evolution according to the Planck $\Lambda$CDM best fit cosmology and in purple we show the evolution in a fully ionized universe.}\label{fig:plasmamass}
\end{figure}

There are always some charged particles in the early Universe and this changes the global propagation properties of photons in cosmology. In particular, the photon gets a mass-squared that is proportional to the number of free electrons in the plasma~\cite{Raffelt:1987im}:
\begin{align}\label{eq:omega_p_app}
    \omega_p^2(z) = \frac{4\pi\alpha_{\rm EM} \, n_e(z)}{m_e}\,,
\end{align}
where $\alpha_{\rm EM}$ is the electromagnetic fine structure constant, $n_e$ is the (free) electron number density and $m_e$ is the electron mass. Photons with energy $E_\gamma < \omega_p(z)$ cannot propagate. This in turn has important implications for the axion decay into photons because by pure kinematics the stars cannot explode into photons unless $m_a > 2 \omega_p(z)$. In Figure~\ref{fig:plasmamass} we show twice the plasma frequency as a function of redshift for the Planck best fit $\Lambda$CDM cosmology (as well as for considering a fully ionized universe). We see that axions with masses $m_a < 2\times 10^{-14}\,{\rm eV}$ cannot decay and thus our constraints are restricted to sufficiently massive axion-like particles. 

The main cosmological implication of the plasma mass is that all the axion stars will not be able to explode until $\omega_p(z) \leq  m_a/2 $. For the range of redshifts $z_{\rm reio}\lesssim z\lesssim 800$ the relation between the redshift of decay and the axion mass is approximately given by:
\begin{align}\label{eq:zdecay}
    z_{\rm decay} \simeq 32 \,\left(\frac{m_a}{10^{-13}\,{\rm eV}}\right)^{2/3}-1\,.
\end{align}
This means that for axion masses in the range $2\times 10^{-14}\lesssim m_a \lesssim 10^{-12}\,{\rm eV}$ all axion stars that are massive enough will explode and convert all their energy into very low energy photons that can subsequently heat the intergalactic medium. For $m_a \gtrsim 10^{-12}\,{\rm eV}$ we will instead expect that axion stars to explode as soon as they form. In this regime, we expect a continuous emission of energy across redshift.

\textit{Effect of Over/Under-densities}: We note that the plasma frequency displayed in Figure~\ref{fig:plasmamass} is the one corresponding to the average electron density in the Universe. However, one could imagine that axion stars are at some point surrounded by either overdense or underdense baryonic environments. First, axion stars do form in the centers of dark matter halos and thus one could at first sight expect them to be surrounded by an overdense baryonic medium~\cite{Fakhouri:2008cn}. However, the halos that host the axion stars that we consider in our study have masses $M_h < 10^5 \,M_\odot $. This in turn means that their potential wells are not deep enough to capture any significant amount of baryonic gas~\cite{Tegmark:1996yt,OLeary:2012gem} and would thus we expect them to be surrounded by the average baryon density. It is also possible that some axion stars form in regions where the baryon density is below the cosmic average. This would in turn mean that one could in principle extend the constraints we discuss to slightly lower axion masses and this has been discussed in the context of light dark photon dark matter in~\cite{Caputo:2020bdy,Witte:2020rvb,Caputo:2020rnx,Garcia:2020qrp}. However, it is important to notice that the explosion mechanism of axion stars is independent on the value of the plasma frequency surrounding the star. The only effect of the plasma frequency would be to potentially delay the explosion itself. Furthermore, noting that the plasma frequency is only mildly sensitive to the baryon density, $\omega_p \propto \sqrt{n_e}$, this means that we do not expect a significant effect from considering the small effect of over or under densities.

%%%%%%%%%%%%%%%%%%%%%%%%%%%%%%%%%%%%%%%%%%%%%%%%%%%%%%%%%%%%%%%%%%%%%%%%%%%%%%%%%%%%
\section{Absorption of photons from axion star explosions in the IGM}\label{sec:absorption}
%%%%%%%%%%%%%%%%%%%%%%%%%%%%%%%%%%%%%%%%%%%%%%%%%%%%%%%%%%%%%%%%%%%%%%%%%%%%%%%%%%%%

Axion star explosions generate a huge number of very low energy photons with $E_\gamma = m_a/2$. For $m_a\lesssim 10^{-8}\,{\rm eV}$ the most efficient absorption process of these photons in the IGM is free-free absorption also known as inverse Bremsstrahlung, $\gamma \, e^-\, p \to e^- p $. The rate at which these photons are absorbed by the plasma is~\cite{Chluba:2015hma}:%see Eq. 29b 
    \begin{align}\label{eq:absorption_1}
        \Gamma_{\rm abs} = n_e \sigma_T  \frac{ \Lambda_{\rm BR}(E_\gamma,z)(1-e^{-E_\gamma/T_b})}{(E_\gamma/T_b)^3}\,,
    \end{align}
where $\sigma_T$ is the Thomson cross section, $T_b$ is the temperature of the electron-baryon fluid and where 
    \begin{align}
    \Lambda_{\rm BR}(E_\gamma,z) = g_{\rm BR} \,\alpha_{\rm EM}\,\frac{n_p}{m_e^3}\sqrt{\frac{2}{3}} 2 \pi ^{3/2} \left(\frac{T_b}{m_e}\right)^{-7/2}\,.
    \end{align}
$g_{\rm BR}$ is the Gaunt factor that for $E_\gamma \ll T_b$ can be approximated by $g_{\rm BR} = \frac{\sqrt{3}}{\pi}\log(2.25 T_b/E_\gamma)$. Plugging in numerical values we find:
\begin{align}\label{eq:absorption}
\Gamma_{\rm abs} &\simeq 2.6\times 10^{-22}\,{\rm eV} \left[\frac{x_e}{2\times 10^{-4}}\right]^2 \nonumber \\
&\times \left[\frac{10^{-13}\,{\rm eV}}{m_a}\right]^{2}\left[\frac{1+z}{21}\right]^{6} \left[\frac{9\,{\rm K}}{T_b}\right]^{3/2} \,,
\end{align}
where here we have normalized the rate to the thermodynamic values of $T_b$ and $x_e$ as expected pre-reionization in $\Lambda$CDM at $z = 20$. We note that the rate has a strong redshift dependence and that it scales as $m_a^{-2}$ and $T_b^{-3/2}$. This means that as the temperature of the baryons raises the free free absorption  becomes less effective. Importantly, it also scales as $x_e^2$ which means that as the Universe becomes ionized the absorption becomes highly efficient.

This rate should be compared with the Hubble expansion $H(z)$ to see if these photons will be absorbed, which is well described around the redshifts of interest by
\begin{align}
        H(z) \simeq 8 \times 10^{-32}\,{\rm eV} \left[\frac{1+z}{21}\right]^{3/2}\,.
\end{align}

\begin{figure}[t]
\centering
\hspace{-0.3cm} \includegraphics[width=0.48\textwidth]{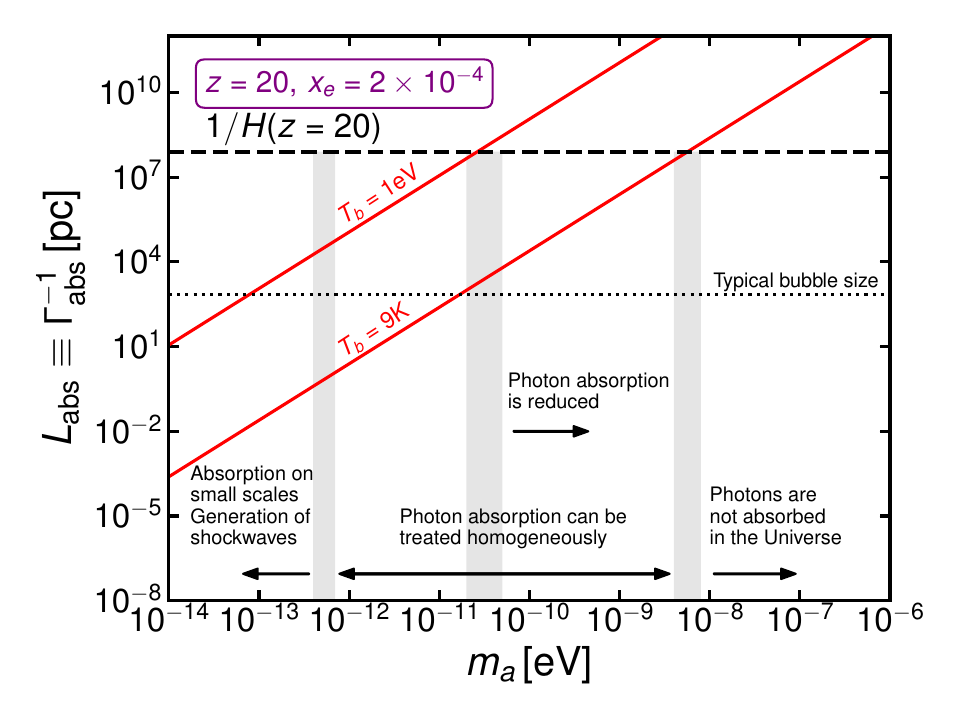}
\vspace{-0.4cm}
\caption{Typical length scale over which photons from axion star decays are absorbed in the IGM via inverse Bremsstrahlung on free electron-proton pairs with a plasma at two different temperatures, $T_b = 9\,{\rm K},\,1\,{\rm eV}$. We show it for the reference value of $z = 20$ and with a pre-reionization free electron fraction $x_e = 2\times 10^{-4}$. We can appreciate three regimes: at $m_a \gtrsim 10^{-8}\,{\rm eV}$ the photons have mean free paths larger than the size of the Universe and therefore there are no cosmological signatures, for $m_a \lesssim 5\times 10^{-11}\,{\rm eV}$ the length-scale of absorption is always smaller than $1/H$ which mean that all of the photons will be absorbed even if $T_b \sim 1\,{\rm eV}$ which is approximately the temperature of an ionized IGM. Finally, for $m_a \lesssim 5\times 10^{-13}\,{\rm eV}$ the photons from axion stars are absorbed on very small lengthscales and thus generates an intense shock-wave which in turn generates a sort of patchy reionization.}\label{fig:absorption}
\end{figure}

Figure~\ref{fig:absorption} explicitly shows the absorption lengthscale of photons produced from axion star decays for two characteristic baryon temperatures, $T_b = 9\,{\rm K}$, and $T_b =1\,{\rm eV}$, which correspond to a cool IGM and one where the IGM is would be fully ionized. From this figure we notice three distinctive regions of parameter space:
\begin{enumerate}
    \item For $m_a \gtrsim 10^{-8}\,{\rm eV}$ the photons have mean free paths larger than the size of the observable Universe and thus axion star explosions will not lead to relevant cosmological implications. 
    \item For $5\times 10^{-13}\,{\rm eV}\lesssim  m_a \lesssim 10^{-8}\,{\rm eV}$ the photons produced from axion star explosions will be absorbed by the IGM on length scales larger than the one it will take for this region to reach $T_b\sim 1\,{\rm eV}$, which is:
    \begin{align}
    & L_{T_b} = \sqrt[3]{M_S/(n_b 3 T_b)} \\ 
    \qquad&\simeq  0.4 \,{\rm kpc}  \left(\frac{M_S}{10^{-4}\,M_\odot} \right)^{1/3} \left(\frac{T_b}{\rm eV} \right)^{1/3} \left(\frac{21}{1+z}\right) \,.  \nonumber
\end{align}
    That means that the absorption of photons in this region of parameter space can be seen as homogeneous, so it will take several nearby axion star explosions to heat up the IGM. This scale is labelled as typical bubble size in Figure~\ref{fig:absorption}.

    \item  For $ m_a\lesssim 5\times 10^{-13}\,{\rm eV}$ the photons are absorbed on very small lengthscales. Since a large amount of energy is released per axion star decay $E\sim M_S \sim 10^{-4}\,M_\odot$ the injection of these photons will lead to shockwaves in the IGM very similar to supernova remnants. The lower limit on $m_a$ of $2\times 10^{-14}\,{\rm eV}$ arises due to the photon plasma mass in the Universe. 
\end{enumerate}

It is important to note that these considerations have been made with $x_e\sim 2\times 10^{-4}$ as expected pre-reionization in $\Lambda$CDM. As soon as $x_e$ grows photons from axion star decays will be absorbed faster in the IGM, see Eq.~\eqref{eq:absorption}.

In what follows, in Appendix~\ref{sec:xe_Tb_evolution} we outline our modeling of energy injections for $m_a\gtrsim 5\times 10^{-13}\,{\rm eV}$ and in Appendix~\ref{sec:evolution_patchy} we consider the case of $m_a\lesssim 5\times 10^{-13}\,{\rm eV}$.

%%%%%%%%%%%%%%%%%%%%%%%%%%%%%%%%%%%%%%%%%%%%%%%%%%%%%%%%%%%%%%%%%%%%%%%%%%%%%%%%%%%%
\section{Continuous Axion Star Mergers and Explosions: Homogeneous IGM Heating and Reionization}\label{sec:xe_Tb_evolution}
%%%%%%%%%%%%%%%%%%%%%%%%%%%%%%%%%%%%%%%%%%%%%%%%%%%%%%%%%%%%%%%%%%%%%%%%%%%%%%%%%%%%

For $m_a\gtrsim 5\times 10^{-13}\,{\rm eV}$ axion stars explode into photons that are absorbed across lengthscales which are larger than the typical interseparation of axion stars. In this regime, the injection of energy can be seen as homogeneous. As discussed in the main text, we track the evolution of the baryon temperature including the new heating source from axion star explosions and also account for Compton and adiabatic cooling. We also consider for the net baryon cooling generated by collisional ionizations $e\,H\to 2e \, p $, as well as collisional excitations $e\,H \to e H^*\to e H \gamma$. We track the free electron fraction using the effective 3-level atom and follow the evolution of protons ($p$) and ${\rm HeII}\equiv {\rm He}^+$. $x_i \equiv  n_i/n_H$ so that $x_e \equiv n_e/n_H = x_{\rm p} + x_{\rm HeII}$. The evolution equations that describe them are:

\begin{widetext}
\begin{subequations}\label{eq:evolutioneqs}
    \begin{align}
\!\!\!\! \frac{\text{d}x_{\rm p}}{\text{d}t} &=\mathcal{C}_{\rm H}\left(\beta_{\rm H}(T_{\gamma})(1-x_{\rm
    p})\text{e}^{\frac{-E_{\text{H}, 2s1s}}{T_{\gamma}}} - x_{\rm e}x_{\rm p}n_{\rm H}\alpha_{\rm H}^{(2)}(T_b) + \frac{\text{d}x_{\rm p}}{\text{d}t}\Biggr|_{\rm coll}\right) \,, \label{eq:dXpdz} \\  
\frac{\text{d}x_{\rm HeII}}{\text{d}t} &=\mathcal{C}_{\rm He}\left((f_{\rm He}-x_{\rm HeII})\beta_{\rm He}(T_{\gamma})\text{e}^{\frac{-E_{\text{He}, 2s1s}}{T_{\gamma}}} - x_{\rm HeII}^2n_{\rm H}\alpha_{\rm HeII}^{(2)}(T_b)+\frac{\text{d}x_{\rm HeII}}{\text{d}t}\Biggr|_{\rm coll}\right) \,, \label{eq:dXHedz}\\
    \frac{\text{d}T_{{b}}}{\text{d}z}(1+z) &= 2\,T_{{b}} +\frac{8}{3}\frac{\rho_\gamma \sigma_T}{m_e H}\frac{x_e}{1+f_{\rm He}+x_{\rm e}}(T_{{b}} - T_\gamma) + \frac{2}{3}\frac{1}{n_{\rm H}(1+f_{\rm He}+x_{\rm e})}\left[\frac{\text{d}E}{\text{d}V\text{d}z}\bigg{|}_{\rm coll}-\frac{\text{d}E}{\text{d}V\text{d}z}\bigg{|}_{\rm dep, h}\right]  \,, \label{eq:Tm_ODE}
\end{align}
\end{subequations}

\end{widetext}

which represent the rate equations for the free electron fraction contribution due to Hydrogen, Eq.~\eqref{eq:dXpdz} Helium, Eq.~\eqref{eq:dXHedz}, recombination ($\alpha$), photoionization ($\beta$) and collisional processes (coll). Both recombination and collisional processes depend on the baryon temperature, $T_b$, which is solved via the rate equation, Eq.~\eqref{eq:Tm_ODE}, where the first term is simply adiabatic cooling, the second corresponds to Compton cooling, the third to the net gas cooling generated by possible collisions in the plasma, and the last term is a result of heating by axion star explosions. The energy deposited as heat in the IGM by axion star explosions is explicitly given by:
\begin{align}
    \frac{\text{d}E}{\text{d}V\text{d}z}\bigg{|}_{\rm dep, h} = \Theta[m_a-2\omega_p] \frac{\text{d}f_{\rm DM}^{\rm decay}}{\text{d}z}\rho_{\rm DM} \left[1-e^{-\Gamma_{\rm abs}/H}\right],
\end{align}
where here the $\Theta$ function ensures that no emission is generated if the photon plasma mass blocks the decay, the last factor is an efficiency factor that takes into account that photons may not be absorbed if their mean free path is large, and finally the fraction of dark matter decaying into photons from axion star mergers is obtained from~\cite{Du:2023jxh} and approximately reads as:
\begin{align}
    \frac{\text{d}f_{\rm DM}^{\rm decay}}{\text{d}z} = \frac{M_S^{\rm decay}}{\rho_{\rm DM}} \frac{\text{d}n_{\rm merge}}{\text{d}z}\,.
\end{align}

The relevant energy transitions in the effective-three level model correspond to the ground state (1s), the second level with two quantum states (2s and 2p), and the third level (c), which denotes the continuum. Direct recombination and photoionization are prohibited, and the Lyman-$\alpha$ decay channel is heavily dampened by the optically thick plasma during the recombination epoch. Therefore, the only possible route to recombination (the ground state, 1s) is to go via the 2s$\xrightarrow[]{}$1s $\gamma\gamma$-photon decay channel. This occurs at a much slower rate compared to Lyman-$\alpha$, $\Lambda_{\rm H, 2s1s}= 8.22458$~s$^{-1}$ which has a transitional energy, $E_{\text{H}, 2s1s}=E_{\text{H}, 2p1s}=10.2$ eV, for Hydrogen, and $\Lambda_{\rm He, 2s1s}=51.3$~s~$^{-1}$ with an energy transition of $E_{\text{He}, 2s1s}=20.62$ eV, for Helium. This creates an overall effective decay rate for energy transitions between 2$\xrightarrow[]{}$1, where $\gamma\gamma$-decay channel dominates at earlier times. This effective decay rate is encapsulated by the Peebles C-factor, which scales the rate equations in Eq.~\eqref{eq:dXpdz} and \eqref{eq:dXHedz} due to these physical effects. The C-factor for Hydrogen, $\mathcal{C}_{\rm H}$, and Helium, $\mathcal{C}_{\rm He}$, are given by, 

\begin{equation}\label{CH}
    \mathcal{C}_{\rm H} = \frac{1+K_{\rm H}\Lambda_{\rm H, 2s1s}(1-x_{p})n_{H}}{1+K_{\rm H}(\Lambda_{\rm H, 2s1s}+\beta_{\rm H})(1-x_{p})n_{H}},
\end{equation}
\begin{equation}\label{CHe}
    \mathcal{C}_{\rm He} = \frac{\left(1 + K_{\rm HeI} \Lambda_{\rm He, 2s1s} n_{\rm H}
  (f_{\rm He}-x_{\rm He II}){\rm e}^{-E_{2s,2p}/T_\gamma}\right)}{\left(1+K_{\rm HeI}
  (\Lambda_{\rm He, 2s1s} + \beta_{\rm HeI}) n_{\rm H} (f_{\rm He}-x_{\rm He II})
  {\rm e}^{-E_{2s,2p}}/T_\gamma\right)},
\end{equation}
%--------------------------------------------------------

respectively, where $K_{\rm H}=\lambda^3_{\text{H}, \text{Ly}\alpha}/8\pi{H}(z)$, $\lambda^3_{\text{H}, \text{Ly}\alpha}= 121.56$ nm, and for Helium, $E_{2s,2p}=0.60$ eV and $K_{\rm HeI} = \lambda^3_{\text{HeI,\;Ly}\alpha}/8\pi {H}(z)$, where $\lambda_{\text{HeI,\;Ly}\alpha}=58.43$ nm \cite{Seager:1999bc}. The recombination rates for Hydrogen, $\alpha_{\rm H}(T_{\rm b})$ and Helium, $\alpha_{\rm HeI}(T_b)$ in Eq.~\eqref{eq:dXpdz} and \eqref{eq:dXHedz}, are taken from, \cite{10.1093/mnras/268.1.109, 1991A&A...251..680P} and \cite{hummer1998recombination}, respectively. 

Assuming local thermal equilibrium and applying the fact the absorption ($\alpha$) and emission ($\beta$) processes are in detailed balance, the photoionization rates are given by $\beta_{\rm H}(T_\gamma)=(2\pi m_{\rm e} T_\gamma)^{1.5}{\alpha_{\rm H}(T_\gamma) }\text{e}^{-\frac{E_{\text{H}, 2s,c}}{T_\gamma}}$ for Hydrogen and $\beta_{\rm HeI}(T_\gamma)=(2\pi m_{\rm e} T_\gamma)^{1.5}{\alpha_{\rm HeI}(T_\gamma) }\text{e}^{-\frac{E_{\text{HeI},2s,c}}{T_\gamma}}$ for Helium. Here $E_{\text{H}, 2s,c}=E_{\text{H}, 1s,c}-E_{\text{H}, \text{Ly}\alpha}=3.39$ eV and $E_{\text{HeI}, 2s,c}=E_{\text{HeI},1s,c}-E_{\text{HeI\;Ly}\alpha}=3.98$ eV. 

Finally, we account for collisional ionizations, i.e. $e \,H\to 2e \,p$ type processes. The change on the number of free ions is given by 
%-------------------------------------------------------
\begin{subequations}\label{eq:collisional_excitation_rate_equation}
\begin{align}%see eq. 8.180 of  2010gfe..book.....M
    \frac{\text{d}x_{\rm p}}{\text{d}t}\Biggr|_{\rm coll} &= \langle\sigma v\rangle_{\rm H}x_{\rm e}n_{\rm H}(1-x_{\rm p})\,,\\
    \frac{\text{d}x_{\rm HeII}}{\text{d}t}\Biggr|_{\rm coll} &=\langle\sigma v\rangle_{\rm HeII}x_{\rm e}n_{\rm H}(f_{\rm He}-x_{\rm HeII})\,\,,
\end{align}
\end{subequations}
%--------------------------------------------------------
where $f_{\rm He}=n_{\rm He}/n_{\rm H}$, and where $\langle\sigma v\rangle_{\rm H}$ and Helium, $\langle\sigma v\rangle_{\rm HeII}$, are the collisional ionization rates for Hydrogen and Helium taken from \cite{doi:10.1063/1.555700}. {The net cooling generated by collisional ionizations and excitations is given by 
%-------------------------------------------
\begin{align}
\frac{\text{d}E}{\text{d}V\text{d}z}\bigg{|}_{\rm coll} &= \frac{n_H}{H} \left[E_{\rm H} \times \frac{\text{d}x_{\rm p}}{\text{d}t}\Biggr|_{\rm coll}  +  E_{\rm HeII} \times \frac{\text{d}x_{\rm HeII}}{\text{d}t}\Biggr|_{\rm coll} \right] \nonumber \\
&+ \frac{\Lambda|_{\rm HI-coll}^{\rm ex}}{H}+ \frac{\Lambda|_{\rm HeII-coll}^{\rm ex}}{H}\,,
\end{align}

%-------------------------------------------
where $E_{\rm H} = 13.6\,{\rm eV}$ and $E_{\rm HeII} = 24.6\,{\rm eV}$. The first two terms arise due to cooling generated by ionization while the last term corresponds to cooling generated by collisional excitations. Namely, by processes of the type $e H \to e H^* \to e H \gamma $, this is, collisions that are not able to ionize the neutral atom can nevertheless excite it and it will subsequently decay back to the ground state by emitting a photon. Since these photons are not absorbed by the gas the whole process cools it. We take these rates from~\cite{Bolton:2006pc} and which read: 
\begin{align}
 \Lambda|_{\rm HI-coll} & =7.50 \times 10^{-19} \,{\rm erg\,s^{-1}\,cm^{-3}} \, e^{-\frac{118348\, {\rm K}}{T_b}} \\
&\times [1+\sqrt{T_b/10^5\,{\rm K}}]^{-1}\, \left[\frac{n_H}{\rm cm^{-3}}\right]^2 x_e (1-x_p)  \nonumber \,,\\
\Lambda|_{\rm HeII-coll} & = 5.54 \times 10^{-17}  \,{\rm erg\,s^{-1}\,cm^{-3}}  e^{-\frac{473638 \,{\rm K}}{T_b}}  \\
&\times (T_b/{\rm K})^{-0.397}  [1+\sqrt{T_b/10^5\,{\rm K}}]^{-1} \nonumber \\
&\times \left[\frac{n_H}{\rm cm^{-3}}\right]^2 x_e (f_{\rm He}-x_{\rm He-II}) \nonumber \,.
\end{align}
}

\begin{figure}[t]
\centering
\includegraphics[width=0.48\textwidth]{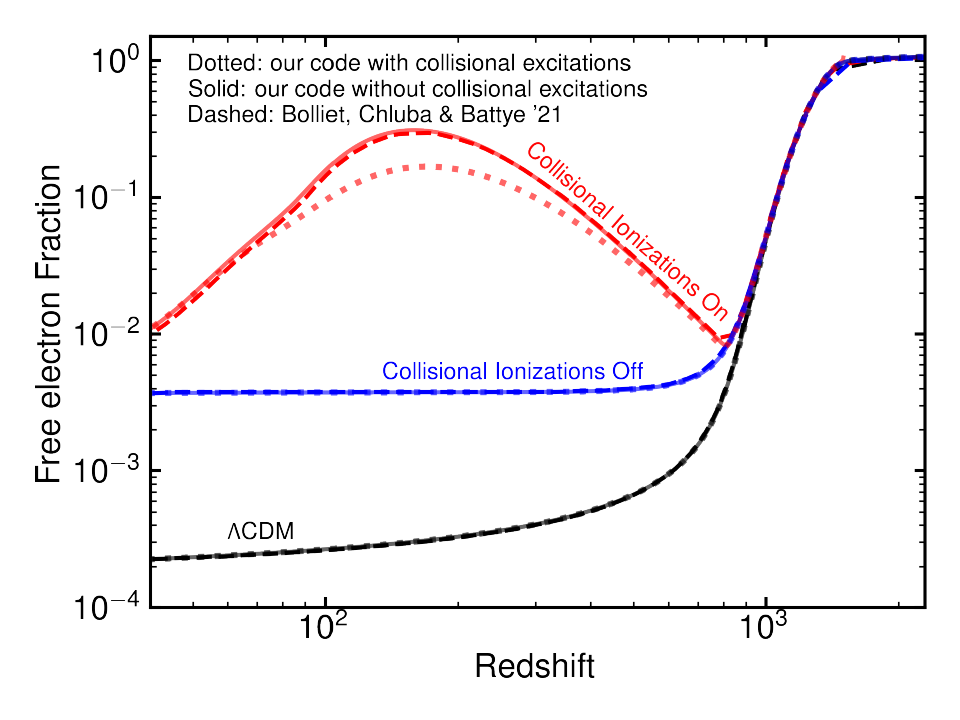}
\vspace{-0.6cm}
\caption{Here we compare the free electron fraction evolution in a scenario of very light decaying dark matter that decays homogeneously and following a typical exponential decay law~\cite{Bolliet:2020ofj}. For the comparison we have used the very same values as in Figure 15 of~\cite{Bolliet:2020ofj}. We can appreciate the excellent agreement when collisional excitations are not considered, as is the case of~\cite{Bolliet:2020ofj}. When collisional excitation cooling is considered, the effect on $x_e$ is somewhat reduced.} \label{fig:xe_comp}
\end{figure}

To cross check our code we considered also the decaying dark matter scenario studied in Ref.~\cite{Bolliet:2020ofj}. There the authors consider a scenario with dark matter decaying into very low energy photons as we do. In their case the decay is modelled by a simple exponential law and in Figure~\ref{fig:xe_comp} we show the excellent agreement for the evolution of the free electron fraction when we use their model of dark matter. {In this reference, the authors did not include the cooling from collisional excitations and it is in that case where we find good agreement. In our study of axion star explosions we do include it since it has a relevant impact (as shown also in Figure~\ref{fig:xe_comp}). }

\textit{Thomson CMB Optical Depth:} Finally, after solving Equations~\eqref{eq:evolutioneqs} we then calculate several integrated quantities. In particular, the Thomson optical depth to recombination which reads:
%-------------------------------------------------%
\begin{equation}\label{doptical_depth_deta}
    \tau = \sigma_{\text{T}}n^{\textit{free}}_{\text{e},0}\int_{0}^{z_{\rm max}} x_e(z)\frac{(1+z)^{2}}{H(z)}\rm{d}z\,.
\end{equation}
%-------------------------------------------------%
In practice we choose $z_{\rm max} = 50$ and since we know the Universe was reionized by $z = 6$ we use $x_e = 1$ for $z \leq 6$ irrespective of the reionization evolution that axion star explosions lead to. Solutions for this reionization evolution are shown in Figure~\ref{fig:xe_Tb_cases}.

\textit{Spectral Distortions:} Additional heating of the IGM due to axion star explosions can generate a $y$-type distortion of the CMB at the redshifts of interest. The maximum amount of energy that could go into the CMB from axion star explosions is given by,
%-------------------------------------------------%
\begin{align}\label{eq::yQdot}
    y &\simeq \frac{1}{4}\frac{\delta \rho_\gamma}{\rho_\gamma} = \frac{1}{4}\int_{\rm z_{\rm min}}^{\rm z_{\rm max}}  \frac{\text{d}Q/\text{d}z}{\rho_\gamma} \text{d}z  \\
    &=\frac{1}{4}\int_{\rm z_{\rm min}}^{\rm z_{\rm max}}  \frac{1}{\rho_\gamma}\frac{\text{d}E}{\text{d}V\text{d}z}\bigg{|}_{\rm dep, h}\text{d}z\,,\nonumber
\end{align}
%-------------------------------------------------%
where $\text{d}Q/\text{d}z$ is the heating rate and $\rho_\gamma$ is the energy density of CMB photons. In Eq.~\eqref{eq::yQdot}, we have chosen to integrate from $z_{\rm min} = 0$ to $z_{\rm max} = 400$, given $z = 400$ is a high enough redshift before axion stars have started to decay. With this expression, we have found that for the parameter space that is currently excluded by Planck constraints on $\tau$ the $y$ distortion is much smaller than the one currently tested by COBE/FIRAS. In particular, for the red circle benchmark point in  Figure~\ref{fig:constraints} we find $|y| = 3\times 10^{-7}$. This is two orders of magnitude smaller than the current sensitivity. For the diamond point in purple that can be tested by 21cm observations, we find $|y| = 1\times 10^{-8}$, which could be within the sensitivity of future CMB observations.

%%%%%%%%%%%%%%%%%%%%%%%%%%%%%%%%%%%%%%%%%%%%%%%%%%%%%%%%%%%%%%%%%%%%%%%%%%%%%%%%%%%%
\section{Patchy Reionization: Growth of Bubbles and Initial Ionization of the Plasma}\label{sec:evolution_patchy}
%%%%%%%%%%%%%%%%%%%%%%%%%%%%%%%%%%%%%%%%%%%%%%%%%%%%%%%%%%%%%%%%%%%%%%%%%%%%%%%%%%%%

For $m_a \lesssim 5\times 10^{-13}\,{\rm eV}$ photons generated by axion star explosions in the early Universe are absorbed on very small lengthscales (see Figure~\ref{fig:absorption}). Axion star explosions release a huge amount of energy $M_S \sim 10^{-4}M_\odot$ (see Figure~\ref{fig:MCrit_Core-Halo}) and will generate a shockwave very similar to those generated by supernova explosions that would be very violent and fully reionize the IGM around it. In this appendix we describe how we track the evolution of the bubbles generated by these explosions and calculate the volume average ionization fraction of the Universe. This scenario leads to patchy ionization.

The expansion of post explosionary shock-waves in a dense medium was first studied in the context of nuclear bombs \cite{taylor,sedov}.  It was later realised that this could also be applied to astrophysical environments such as around supernova explosions~\cite{Avedisova} and the equations were subsequently developed to include the internal and external pressure changes, the heating of the interior of the bubble \cite{1977ApJ...218..377W}, energy lost to the shockwave due to the emission of radiation~\cite{1975ApJ...200L.107C}, the self gravity of the shell and the expansion of the Universe \cite{Ostriker:1988zz,1993ApJ...417...54T,1996ApJ...465..548V}.

In order to model the behaviour of the expanding bubble with radius $R$ and  enclosed mass $M$, we use the following two coupled equations (see, e.g., \cite{1993ApJ...417...54T,Pieri:2006bd,Blas:2020nbs}):
\begin{subequations}\label{eq:evolutionbubble}
\begin{align}
\label{eq:rdotdot}
\!\!\!\!  \ddot R&={8\pi G\,p\over\Omega_bH^2R}-{3\over R}(\dot R-HR)^2
-{\Omega_m H^2R\over2}-{GM\over R^2}\,,\\
\label{eq:pdot}
\!\!\!\!  \dot p&={L_{\rm tot}\over2\pi R^3}-{5\dot Rp\over R}\,,
\end{align}
\end{subequations}

\noindent where a dot represents derivative w.r.t time, $\Omega_m$ is the total density parameter, $\Omega_b$ the baryon density parameter, and $H$ is the Hubble expansion rate. $L_{\rm tot}$ represents the total luminosity and $p$ is the bubble pressure resulting from this luminosity. 

The first term in equation~\eqref{eq:rdotdot} represents the driving pressure of the outflow, the second is drag due to accelerating the IGM from velocity $HR$ to velocity $\dot R$, the third is the self gravity of the expanding shell the fourth is the self gravity of the entire halo.
The first term in equation~\eqref{eq:pdot} represents the increase in pressure caused by injection of energy while the second term is the drop in pressure caused by adiabatic expansion. In particular, the total luminosity is given by: 
\begin{align}
L_{\rm tot} = L_{\rm Explosion} - L_{\rm Compton} - L_{\rm Ionization}\,,
\end{align}
where here $L_{\rm Explosion}$ is the luminosity that is generated by the explosion of the axion star into photons, $L_{\rm Compton}$ is the luminosity lost via Compton cooling against the CMB, and $L_{\rm Ionization}$ takes into account the energy lost in ionising the swept IGM by the shockwave. 

Since the timescale for the explosion is very short compared to any other time scale, see Eq.~\eqref{eq:Mdecay}, and the one for absorption as well, see Eq.~\eqref{eq:absorption}, in practice we assume that the blastwave expands freely until the mass contained in the bubble is comparable to the energy ejected by the axion star explosion. At that point we start the integration of Eqs.~\eqref{eq:rdotdot} and~\eqref{eq:pdot} neglecting the explosion luminosity but taking into account the fact that a large pressure has been generated by the explosion. This is similar to what is done for supernova explosions, see Section 8.6.1 of~\cite{2010gfe..book.....M}, and the starting radius, the velocity and the pressure read as follows:
\begin{subequations}\label{eq:initialconditions}
\begin{align}%see eq. 8.180 of  2010gfe..book.....M
    R_0 &= 0.7\,{\rm pc} \left(\frac{M_S}{10^{-4}\,M_{\odot}}\right)^{1/3}\,,\\
    \dot R_0 &=0.5\, R_0 \, , \\ 
    p_0 &= \frac{M_S}{2\pi R_0^3}\,.
\end{align}
\end{subequations}
We thus assume that the pressure density is at that point essentially given by the energy output in the axion star explosion in the volume filled by the bubble and that the bubble is moving relativistically at a speed of $v = 0.5\,c$. From the initial conditions in Eq.~\eqref{eq:initialconditions} we can then solve Eqs.~\eqref{eq:evolutionbubble} with $L_{\rm Explosion} = 0$. The last thing needed is the other luminosities which can be written as~\cite{1993ApJ...417...54T}:
\begin{align}
    L_{\rm Compton} = \frac{2\pi^3}{45} \frac{\sigma_T}{m_e} T_\gamma^4\,p R^3\,,
\end{align}
where $T_\gamma$ is the CMB temperature at a given redshift, and
\begin{align}
    L_{\rm Ionization} = f_m \,n_b\, E_H4\pi^2 R^2 (\dot{R}-H R)\,,
\end{align}
where here $n_b$ is the background baryon number density, $E_H\simeq 13.6\,{\rm eV}$ is the energy it takes to ionize a hydrogen atom, and $f_m$ is the fraction of the baryonic mass kept in the interior of the bubble which by construction is $f_m\ll 1$. In our calculations, for concreteness we take $f_m = 0.1$ but we have checked that smaller values yield very similar results. Finally, we stop the integration when the pressure of the interior of the bubble is $p \simeq 2\,T_{\rm crit} n_b $ with $T_{\rm crit} \simeq 15000\,{\rm K}$ which roughly corresponds to the temperature at which the IGM becomes is fully ionized. For smaller pressures the bubble will not be able to ionize the swept IGM and the radius at this point will tell us the region that has been ionized by the axion star explosion.

\begin{figure}[t]
\centering
\hspace{-0.3cm} \includegraphics[width=0.48\textwidth]{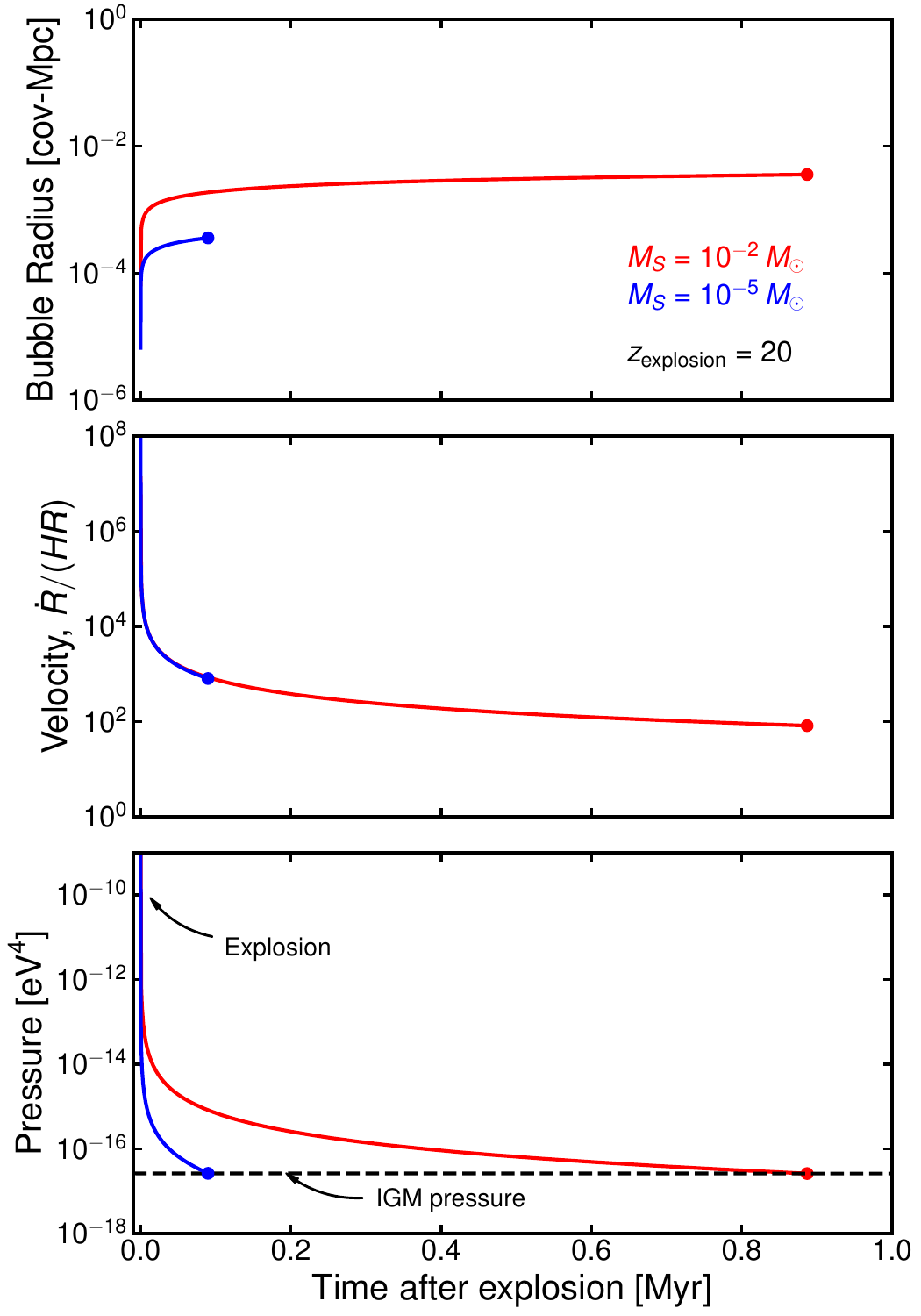}
\caption{Time evolution of the shockwave resulting from an axion star explosion into photons that are readily absorbed in the IGM. Top to bottom we show the comoving radius, the velocity normalised to the Hubble rate, and the pressure inside the bubble as a function of time. }\label{fig:bubbleevolution}
\end{figure}

Figure~\ref{fig:bubbleevolution} shows the numerical solution to Eqs.~\eqref{eq:evolutionbubble} for one energy injection example. For concreteness, we show the result for $z_{\rm decay} = 20$ and for two different axion star masses. There we can clearly see the steep decrease of the pressure as a result of the increase in the bubble radius. We note that we do not find any significant dependence of the results with respect to the redshift of injection. 

By numerically solving for all the axion-star masses of interest we find that the comoving size that the bubbles of ionized material reach is: 
\begin{align}\label{eq:FinalRadius}
    R_{\rm bubble-final}^{\rm comoving} \simeq 0.4\,{\rm kpc} \left(\frac{M_S}{8\times 10^{-5}\,M_\odot} \right)^{1/3}\,.
\end{align}
We have also found that the characteristic time it takes the bubble to reach such a radius is much shorter than the age of the Universe and is approximately given by:
\begin{align}
    t_{\rm bubble} \simeq 0.2\,{\rm Myr} \left(\frac{M_S}{8\times 10^{-5}\,M_\odot}\right)^{1/3}\,.
\end{align}

Eq.~\eqref{eq:FinalRadius} is very useful as it tells us the characteristic size of ionized bubbles in the Universe. This, together with the number density of such axions stars will allow us to understand the global reionization driven by this process. 

Once the bubble pressure reaches the IGM one, the bubbles will stuck. From this point onward the bubble will simply experience Hubble expansion and we can treat the free electron fraction and its temperature following the standard equations for recombination/ionization. In particular, we use the very same equations as in Eq.~\eqref{eq:evolutioneqs} but without a heating term, namely, with $\text{d}E/(\text{d}V\text{d}z)|_{\rm dep,h} = 0$. 

We then solve the set of equations~\eqref{eq:evolutioneqs} starting with $x_e = 1$ and $T_b =1\,{\rm eV}$ as expected from the final state of the bubble. What happens from then onwards is essentially that the bubbles cool due to Compton cooling and this leads to recombination which in turn reduces $x_e$. We simply then calculate the volume average free electron density by taking this result and multiplying it by  $n_S^{\rm crit} \times R^3$ where we obtain $n_S^{\rm crit}$ from~\cite{Du:2023jxh} and $R$ is given by Eq.~\eqref{eq:FinalRadius}. An example of the evolution of this volume averaged ionization fraction can be seen in Figure~\ref{fig:xe_Tb_cases} in dashed blue. We can clearly appreciate the downward trend of $x_e$ as generated by Compton cooling.

%%%%%%%%%%%%%%%%%%%%%%%%%%%%%%%%%%%%%%%%%%%%%%%%%%%%%%%%%%%%%%%%%%%%%
%%%%%%%%%%%%%%%%%%%%%%%%%%%%%%%%%%%%%%%%%%%%%%%%%%%%%%%%%%%%%%%%%%%%%
\bibliography{biblio}
%%%%%%%%%%%%%%%%%%%%%%%%%%%%%%%%%%%%%%%%%%%%%%%%%%%%%%%%%%%%%%%%%%%%%
%%%%%%%%%%%%%%%%%%%%%%%%%%%%%%%%%%%%%%%%%%%%%%%%%%%%%%%%%%%%%%%%%%%%%

\end{document}